\documentclass[preprintnumbers,superscriptaddress,amsmath,amssymb,prd,preprint,nofootinbib]{revtex4-2}
\usepackage{graphicx}
\usepackage{amsfonts}
\usepackage[colorlinks,linkcolor=red,anchorcolor=blue,citecolor=green]{hyperref}
\usepackage{subfigure}
\usepackage{float}
\usepackage{enumerate}
\newcommand\diff{\mathrm{d}}

\begin{document}
\title{
    Viscous cosmology in $f(T)$ gravity
}
\author{Jing Yang}
\author{Rui-Hui Lin}
\email[]{linrh@shnu.edu.cn}
\author{Xiang-Hua Zhai}
\email[]{zhaixh@shnu.edu.cn}
\affiliation{Division of Mathematics and Theoretical Physics, Shanghai Normal University, 100 Guilin Road, Shanghai 200234, China}

\begin{abstract}
    We propose a new model for the viscosity of cosmic matters,
    which can be applied to different epochs of the universe.
    Using this model, we include the bulk viscosities as practical corrections to the perfect fluid models of the baryonic and dark matters since
    the material fluids in the real world may have viscosities due to thermodynamics.
    Such inclusion is put to the test within the framework of $f(T)$ gravity that is proved to be successful in describing the cosmic acceleration,
    where $T$ denotes the torsion scalar.
    We perform an observational fit to our model and constrain the cosmological and model parameters by using various latest cosmological datasets.
    Based on the fitting result, we discuss several cosmological implications including the dissipation of matters,
    the evolutionary history of the universe, $f(T)$ modification as an effective dark energy, and the Hubble tension problem.
    The corresponding findings are
    (i) The late time dissipation will make the density parameters of the matters vanish in the finite future.
    Moreover, the density ratio between the baryonic and dark matters will change over time.
    (ii) The radiation dominating era, matter dominating era and the accelerating era can be recovered
    and the model can successfully describe the known history of the universe.
    (iii) The $f(T)$ modification is the main drive of the acceleration expansion and currently mimics a phantom-like dark energy.
    But the universe will eventually enter a de Sitter expansion phase.
    (iv) The Hubble tension between local and global observations can be significantly alleviated in our model.
\end{abstract}
\maketitle
\newpage
\section{Introduction}
\label{intro}
Cosmological and astrophysical observations including the Type-Ia supernovae (SNIa), cosmic
microwave background radiation (CMB), baryon acoustic oscillations (BAOs) and
large-scale structure (LSS) suggest that our universe is now in an acceleration phase of expansion.
One of the straightforward ways to explain this acceleration
is to introduce the cosmological constant $\Lambda$ back to play the role of the so-called dark energy
and construct the $\Lambda$ cold dark matter ($\Lambda$CDM) model.
This model proves to be quite accurate and successful.
However,
from the theoretical point of view, the cosmological constant $\Lambda$ lacks a physical origin and
its observed value is too small to be explained by any fundamental theories.
Moreover, this simple model also suffers from the cosmic coincident problem, i.e.,
the densities of matter and dark energy of the universe are of the same order.
To alleviate these problems, various dark energy models have been proposed and studied
(for comprehensive reviews, see, e.g., Refs. \cite{Copeland:2006wr,Bahamonde:2017ize,Frusciante:2019xia,Ishak:2018his}).
Another angle to address these problems is to consider modifications of gravity and
see General Relativity (GR) as an approximation to the real theory of gravity yet to be discovered.
One of the simplest schemes to modify Einstein's gravity is the $f(R)$ gravity,
where the curvature scalar $R$ in the gravitational Lagrangian is replaced by a function $f$ of $R$
(see, e.g., Refs. \cite{DeFelice:2010aj,Sotiriou:2008rp,Capozziello:2011et,Nojiri:2010wj} for extensive reviews).
On the other hand, based on an equivalent description of Einstein's gravity, known as the
Teleparallel Equivalent of General Relativity (TEGR) \cite{Aldrovandi:2013wha,Maluf2013}
where the torsion scalar $T$ takes the place of the curvature scalar $R$,
the $f(T)$ modification of gravity is proposed and widely studied
(see, e.g., Refs. \cite{PhysRevD.79.124019,PhysRevD.81.127301,Cai:2015emx,Nojiri:2017ncd}).

In most studies of cosmology, may that be within the framework of GR or modified gravities,
the universe media are usually simplified as perfect fluids.
From the hydrodynamic point of view, a less ideal fluid may involve the dissipative effects that lead to a deviation from the thermal equilibrium.
Such dissipative processes in the cosmic expansion include heat conduction, bulk viscosity, and shear viscosity,
which have been carefully studied (see, e.g., Refs. \cite{Eckart:1940te,Israel:1976tn,Hiscock:1991sp}).
Due to the homogeneous and isotropic assumption of the cosmic background,
considering the bulk viscosity as the only relevant dissipative phenomenon in cosmology is appropriate.
Pioneer studies to include bulk viscosities in cosmology follow conceptually different trains of thought,
partly such as
\begin{itemize}
    \item[$\bullet$] \textbf{Viscosity as a sole drive of acceleration}:
          Either some part or the unity of the universe medium has viscosity which is the sole mechanism to drive the cosmic acceleration
          \cite{Ren:2006en,Meng:2005jy,Cataldo:2005qh,Brevik:2005bj,PhysRevD.75.043521,PhysRevD.76.103516,PhysRevD.79.103521,Avelino:2010pb,PhysRevD.84.103508,Gagnon:2011id,Fabris:2011wk}.
          This idea utilizes the fact that the bulk viscosity of the fluid may act as an effective pressure, possibly negative,
          to recover the local thermodynamic equilibrium \cite{OKUMURA2003207},
          and explores the possibility that this effective pressure may be the culprit of the accelerated expansion of the universe.
          Such a consideration can be seen as one of the approaches to a unifying model of (dark) matter and dark energy.
    \item[$\bullet$] \textbf{Dust matter and viscous dark sector}:
          The known contents of the universe are still without viscosity,
          while the dark sector that has already the mechanism to drive the acceleration may be viscous
          \cite{Hu:2005fu,Feng:2009jr,Setare:2010zz,PhysRevD.82.063507,Montiel:2011gw,Wang:2017klo,Brevik:2017msy,daSilva:2020mvk}.
          Since the properties of dark energy or dark matter or the unification of both are still not very conclusive,
          this idea puts more physical and realistic considerations into the model of dark sector.
    \item[$\bullet$] \textbf{Viscous matter and effective dark energy}:
          The material contents may have viscosities,
          while the accelerated expansion is of geometric or gravitational origin, e.g., cosmological constant or modified gravities
          \cite{PhysRevD.86.083501,Brevik:2012bc,Singh:2014bha,Velten:2013qna,Brevik:2017msy,Hu:2020xus,Arora:2020lsr,Sharif:2013tny,Jawad:2016omn,DavoodSadatian:2019pvq,Gadbail:2021fjf,Arora:2021tuh}.
          This idea sees the viscosities to be the physical properties that may be possessed by any material fluid in the real world
          and considers the viscosity as one of the realistic corrections of imperfectness to the perfect fluid model.
          The inclusion of viscosities in the fluid models is put to the test within the various frameworks of successful cosmological models.
\end{itemize}

In this work, we follow the last route and intend to include the viscosities in the cosmological context of the $f(T)$ gravity.
In Refs. \cite{Sharif:2013tny,Jawad:2016omn,DavoodSadatian:2019pvq},
the authors have analyzed the cosmology in $f(T)$ gravity where the universe is filled with a single viscous fluid
by considering specific evolutionary functions given by $H(t)\propto 1/t^\lambda$,
where $H$ is the Hubble parameter, $t$ is the cosmic time and $\lambda$ is a constant parameter.
Moreover, $f(T)$ gravity has also been used to reconstruct the generalized Chaplygin gas model with viscosity \cite{Saha:2022ian}.
In the current work, to accommodate some of the existing late time viscous models,
we propose a simple unifying bulk viscous model of the matters that can be applied to different epochs of the universe.
Using this new model, we intend to study the evolutionary history of the universe with bulk viscous matter in $f(T)$ gravity,
where different cosmic contents, i.e.,
radiation and viscous baryonic and dark matters, are included.
The motivation of this consideration is twofold.
Firstly, we can model the whole evolutionary history of the universe including the radiation dominating era,
matter dominating era and the acceleration era.
As such, we can include the rather stringent datasets of CMB and BAO in addition to the late time SNIa datasets
and various estimations of $H$ dependence on the redshift $z$ (denoted as Hz in the following) in the parameter fitting.
Secondly, the physics before the last scattering can be different from the standard model
not only for the modification of $f(T)$ gravity but also for the bulk viscosity.
This could be significant.
For example, the recently revealed tension between the Hubble constant $H_0$ measured locally and that fitted by the global observation \cite{Freedman:2017yms}
may, among various possibilities, be a result of modified prerecombination physics \cite{PhysRevD.97.103511,Aylor:2018drw,Verde:2019ivm}.

The paper is arranged as follows.
Section \ref{review} contains brief descriptions of the bulk viscous model of matter contents and the $f(T)$ gravity.
In Sect. \ref{treatments}, we use the observational datasets of CMB, BAO, SNIa, and Hz to fit and evaluate the model,
presenting constraints and best-fit values of the cosmological and model parameters.
Some cosmological implications including the dissipation of the matters, the evolutionary history of the universe,
$f(T)$ modification as an effective dark energy, and the Hubble tension problems are presented in Sect. \ref{implication}.
We conclude our study in the last section.

Throughout the paper, we use the natural unit with $c=8\pi G=1$,
where $c$ is the speed of light and $G$ is the gravitational constant.

\section{Overview of the $f(T)$ gravity and the viscous cosmology}
\label{review}

\subsection{Cosmological model with bulk viscous matter}
We assume a Friedmann-Lema\^itre-Robertson-Walker (FLRW) metric for the spatial flat universe, which is given by
\begin{equation}
    \label{FLRW}
    \diff s^2=\diff t^2-a(t)^2\delta_{ij}\diff x^i\diff x^j,
\end{equation}
where $a(t)$ is the cosmic scale factor.

For the matter source we consider an imperfect fluid with viscosity,
where the dissipative effects resulted from microscopic interactions are described by transport coefficients.
Since the universe is treated to be isotropic and expanding,
only bulk viscosity should be relevant and it can be expressed as a first order deviation from the thermodynamic equilibrium.
The energy-momentum tensor for such a fluid is then given by \cite{Eckart:1940te,Israel:1976tn,Hiscock:1991sp,OKUMURA2003207,Brevik:2017msy},
\begin{equation}
    \label{impfemtensor}
    \bar{\mathcal T}_\mu^{\;\nu}=\left( \rho+\bar p \right) u_\mu u^\nu - \bar p\delta_\mu^\nu,
\end{equation}
with
\begin{equation}
    \label{bulkpressure}
    \bar p=p-\zeta\nabla_\mu u^\mu,
\end{equation}
where  $p$ and $\rho$ are the pressure and energy density of the fluid, respectively,
$u^\mu$ is the 4-velocity, and $\zeta$ is the bulk viscosity arising from deviation from the local thermodynamic equilibrium.
The viscosity is then encoded in a viscous pressure term $-\zeta\nabla_\mu u^\mu$ in addition to the usual material pressure.
In the cosmological scenario, this term can be written as $-3\zeta H$, where $H=\dot a/a$ is the Hubble parameter.

The continuous equation $\nabla_\nu\bar{\mathcal T}_\mu^{\;\nu}=0$ is equivalent to
\begin{equation}
    \label{source1}
    \nabla_\nu\mathcal T_\mu^{\;\nu}=\nabla_\nu \left[ \left( \delta_\mu^\nu-u_\mu u^\nu \right)\left( \bar p-p \right) \right],
\end{equation}
where $\mathcal T_\mu^{\;\nu}$ is the energy-momentum tensor of a usual perfect fluid.
For cosmological metric and the assumption $p=0$,
\begin{equation}
    \label{source2}
    \nabla_\nu\mathcal T_0^{\;\nu}=-3H\bar p,
\end{equation}
where the right-hand side can be either seen as a sink of energy if $\bar p$ is positive,
or a source if $\bar p$ is negative.
It reflects that there is energy transference between the kinetic energy of cosmological expansion
and the material fluid due to the bulk viscosity.

The transport coefficient $\zeta$ should depend on the dynamics of the material fluid.
A general model for viscous effects that emerge at late time is proposed and can be written as
\cite{Ren:2005nw,Avelino:2010pb,Singh:2014bha,Brevik:2017msy,Sasidharan:2018bay,Arora:2020lsr}
\begin{equation}
    \label{zetaH}
    \zeta=\zeta_0+\zeta_1 H,
\end{equation}
where both $\zeta_0$ and $\zeta_1$ are constants.
This can be viewed as an agreement with a generalized equation of state (EOS) of a dark fluid \cite{PhysRevD.73.043512,PhysRevD.101.044010} since the pressure now can be written as
\begin{equation}
    \bar p = p(\rho)+w_0 H+w_1 H^2,
\end{equation}
where $w_0=-3\zeta_0$ and $w_1=-3\zeta_1$ are new constant coefficients of EOS.
Another widely considered model of material bulk viscosity is of the form
\cite{Padmanabhan:1987dg,Szydlowski:2006ma,Hipolito-Ricaldi:2009xbk,Brevik:2017msy,Mohan:2017poq,Sasidharan:2018bay}
\begin{equation}
    \label{zetarho}
    \zeta\sim\rho^s.
\end{equation}
For late time universe when matter is dominating and $H^2\sim\rho_m$,
these two forms \eqref{zetaH} and \eqref{zetarho} can reach some common ground for $s=1/2$.

However, for early times of the universe,
both forms will lead to an unbounded viscosity
or a growing Hubble parameter,
which seems to be contrary to our conventional understanding of the evolutionary history of the universe.
Therefore, to cover the early times,
we consider the merits of both forms of the viscosity and propose a simple new form
\begin{equation}
    \label{zeta}
    \zeta=\zeta_0\sqrt\Omega+\zeta_1\Omega H,
\end{equation}
where $\Omega=\rho/(3H^2)$ is the density parameter of the corresponding content with bulk viscosity.
While Eq. \eqref{zeta} preserves the property \eqref{zetarho},
it will reduce to the form \eqref{zetaH}
at the era when the universe is dominated by the corresponding viscous content with $\Omega\sim 1$,
and will be suppressed when the viscous content is not dominating.

In the current work, for simplicity,
we only consider the imperfectness and bulk viscosity of the material contents
including both the baryonic and dark matters,
leaving the relativistic contents to be described by the EOS $w=p/\rho=\frac13$.
In addition, we assume that the viscosities of the baryonic and dark matters have the same coefficients.
That is,
\begin{equation}
    \label{impmatters}
    \begin{split}
        \overline{p_b}=&p_b-3H(\zeta_0\sqrt{\Omega_b}+\zeta_1\Omega_b H),\\
        \overline{p_m}=&p_m-3H(\zeta_0\sqrt{\Omega_m}+\zeta_1\Omega_m H),
    \end{split}
\end{equation}
where the subscripts $b$ and $m$ indicate the baryonic and dark matters, respectively.
One can easily check that
\begin{equation}
    \overline{p_i}=p_i-\zeta_0\sqrt{3\rho_i}-\zeta_1\rho_i,
\end{equation}
where $i=b,m$.
So, the $\zeta_1$ term is in effect a correction to the pressureless EOS $w_i=0$,
and one can assume the material pressure $p_i=0$ without loss of generality.

It is noted here that the viscous effects of the baryonic and dark matters must be considered separately
even it is simply assumed that they have the same viscous coefficients.
Otherwise the viscosity will become another interaction between the baryonic and dark matters other than gravitation.
At late time, this does not seem very important in that the same viscous coefficients
and EOSs make sure both the baryonic and dark matters evolve synchronously.
However, before the last scattering, the propagation of the sound wave of the photon-baryon fluid
depends closely on the EOS of the baryon matter,
but is not so sensitive to that of the dark matter.
Specifically, around some primordial and initial clumps of matter,
the pressure of the photon-baryon fluid resists the gravitational clustering.
The resulting sound waves propagate at the sound speed $c_s$ given by
\begin{equation}
    \label{soundspeed}
    c_s^2=\frac{\partial p}{\partial \rho}=\frac{\overline{p_b}'+p_\gamma '}{\rho_b'+\rho_\gamma'},
\end{equation}
where the subscript $\gamma$ indicates the photon and the prime denotes derivative
with respect to some common variable, e.g., the scale factor $a$.
The propagations continued until the universe became transparent for photons,
leaving dramatic oscillations CMB anisotropic data we see today,
as well as imprints in the galaxy distributions.
This is the key to our model in the era before the last scattering
and include the CMB and BAO datasets in the following section to perform the observational fit.

\subsection{Viscous cosmology in $f(T)$ gravity}
Although the bulk viscosity may provide a possibly negative pressure, as seen in Eq. \eqref{impmatters},
it can hardly be the main cause of the late time acceleration of the universe.
As we will see in the next section,
the densities and viscosities of the matters are quite stringently constrained by the observation.
The effective pressures from viscosities do not seem to be of the same order of the energy densities
and hence cannot account for the missing dark energy that are around 70 percent of the cosmic contents.
Therefore, the drive of the acceleration is still needed elsewhere.
In this work, we consider the cosmic evolution under the framework of $f(T)$ gravity,
attributing the late time acceleration to the modification of gravitational theory in the teleparallel description.

On the parallelizable spacetime manifold $\mathcal M$ with a metric $g$,
one can generally find a set of tetrad $\{e_A^{\;\mu}\}$ and its dual $\{e^A_{\;\mu}\}$,
such that
\begin{equation}
    e^A_{\;\mu}e_B^{\;\mu}=\delta_B^A,\quad e_A^{\;\mu}e^A_{\;\nu}=\delta_\nu^\mu.
\end{equation}
They are related to the metric tensor via
\begin{equation}
    g_{\mu\nu}=\eta_{AB}e^A_{\;\mu}e^B_{\;\nu},\quad\eta_{AB}=g_{\mu\nu}e_A^{\;\mu}e_B^{\;\nu}.
\end{equation}
Then, the torsion scalar $T$ is defined as
\begin{equation}
    T\equiv T^\alpha_{\;\mu\nu}S_\alpha^{\;\mu\nu},
\end{equation}
where the torsion tenser is
\begin{equation}
    T^{\alpha}_{\hspace{0.5em}\mu\nu}=e_{A}^{\hspace{0.5em}\alpha}\left (\partial _{\mu } e^{A}_{\hspace{0.5em}\nu} - \partial _{\nu }e^{A}_{\hspace{0.5em}\mu} \right )
\end{equation}
and the superpotential is
\begin{equation}
    \label{sp}
    S_\alpha^{\:\beta\gamma}=\frac14 \left( T_\alpha^{\:\beta\gamma}+T^{\gamma\beta}_{\:\;\:\;\alpha}-T^{\beta\gamma}_{\:\;\:\;\alpha} \right)+\frac12 \left( \delta^\beta_\alpha T^{\lambda\gamma}_{\:\;\:\;\lambda}-\delta^\gamma_\alpha T^{\lambda\beta}_{\:\;\:\;\lambda} \right).
\end{equation}
TEGR uses this scalar $T$ as its gravitational Lagrangian,
and the $f(T)$ gravity considers an arbitrary function of $T$ instead, i.e.,
\begin{equation}
    \label{ftlag}
    \mathcal{S}=-\frac12\int|e|f(T)\diff^4 x+\int|e|\mathcal{L}_M\diff^4x,
\end{equation}
where $|e|=\det(e^A_{\alpha})=\sqrt{-g}$ is the determinant of the tetrad $e^A_{\alpha}$,
and $\mathcal{L}_M$ is the Lagrangian of matter.

Varying the above action \eqref{ftlag} with respect to the tetrad $e^A_{\alpha}$,
one can obtain the field equation
\begin{equation}
    \label{fieldeq}
    \frac2{|e|}\partial_\beta \left( |e|S_\sigma^{\:\alpha\beta}e_A^{\:\sigma}f_T \right)+\frac f2 e_A^{\:\alpha}=\mathcal{T}_\beta^\alpha e_A^{\:\beta},
\end{equation}
where $f_T$ denotes $\diff f/\diff T$,
and the energy-momentum tensor $\mathcal{T}_\beta^\alpha$ of matter is given by
\begin{equation}
    \label{emts}
    \frac{\delta(|e|\mathcal{L}_M)}{\delta e^A_{\:\alpha}}=|e|\mathcal{T}_\beta^\alpha e_A^{\:\beta}.
\end{equation}

For the cosmological metric given in Eq. \eqref{FLRW}, one can easily find that $T=-6H^2$.
Then, since the Hubble parameter $H$ is generally a decreasing function of cosmic time,
one would expect that a $f(T)$ model with a modification term more significant at smaller $|T|$ is appropriate
in that it could give explanation to the late time (smaller $H$ and $|T|$) acceleration
and would not mess up the early time evolution.
We therefore choose a concrete $f(T)$ model in the form
\begin{equation}
    \label{fTform}
    f(T)=T+\frac\alpha T,
\end{equation}
where $\alpha$ is the model parameter.
Studies show that the modifications with inverse powers of $T$ may indeed provide the drive of the accelerated expansion
\cite{Feng:2015awr,Lin:2016nvj,Lin:2018xzd}.

Taking the model \eqref{fTform} and the energy-momentum \eqref{impfemtensor} into the field equation \eqref{fieldeq},
one can obtain the modified Friedmann equations
\begin{equation}\label{friedmann}
    \begin{split}
        3H^{2}-\frac{\alpha }{4H^{2}}&=\rho_r+\rho_b+\rho_m,\\
        2\dot{H}\left ( 1+\frac{\alpha }{12H^{4}}  \right )  &=-p_r-\overline{p_b}-\overline{p_m}-\rho_r-\rho_b-\rho_m,
    \end{split}
\end{equation}
where the subscript $r$ denotes the radiation including both the photons and relativistic neutrinos.

\section{Observational fit}
\label{treatments}
In this section, we perform an observational fit to the cosmological model we introduce in the previous section.
In most studies of viscous cosmology, the authors usually focus on a certain epoch of the evolutionary history of the universe.
Since the newly proposed model of viscosity in the current work depends on $\Omega$ instead of $H$ or $\rho$,
we can consider the cosmic evolution as a whole and, more importantly,
utilize the observational datasets of both early and late time eras in a joint fitting procedure.

For the late time SNIa data, we consider the latest Pantheon compilation \cite{Pan-STARRS1:2017jku}
consisting of a total of $1048$ SNIa with the cosmic redshift $z$ ranging from $0.01$ to $2.3$.
We also utilize $57$ data points of Hz \cite{Sharov:2018yvz}.
These two sets of data represent the observations of the acceleration phase of the universe
and provide the late time constraints of the model.

For earlier era, we consider the CMB and BAO datasets.
The CMB temperature power spectrum is sensitive to the matter densities,
and it also measures precisely the sound horizon,
which is given by
\begin{equation}
    \label{rs}
    r_s^*=\int_{z^*}^\infty \frac{c_s}{H(z)}\diff z,
\end{equation}
where $z^*\sim 1000$ is the cosmic redshift of the last-scattering surface
and $c_s$ is the sound speed of the photon-baryon fluid before the last scattering.
In viscous cosmology, such consideration is only possible when the viscosities of baryonic and dark matters are separated
in that $c_s$ is closely related to the EOS of baryonic matter but not sensitive to that of dark matter.
By using the viscosity model \eqref{impmatters} proposed in the previous section,
we can substitute $c_s$ with Eq. \eqref{soundspeed}.
We will be using the CMB data inferred from Refs. \cite{SDSS:2014iwm,Planck:2018vyg}.
Moreover, the BAO measurement provides a standard ruler to probe the angular diameter distance versus the redshift
by performing a spherical average of their scale measurement.
We will use the BAO data given in Refs. \cite{Beutler:2011hx,Padmanabhan:2012hf,Anderson:2012sa}.

\begin{figure*}[htpb]
    \centering
    \includegraphics[width=\linewidth]{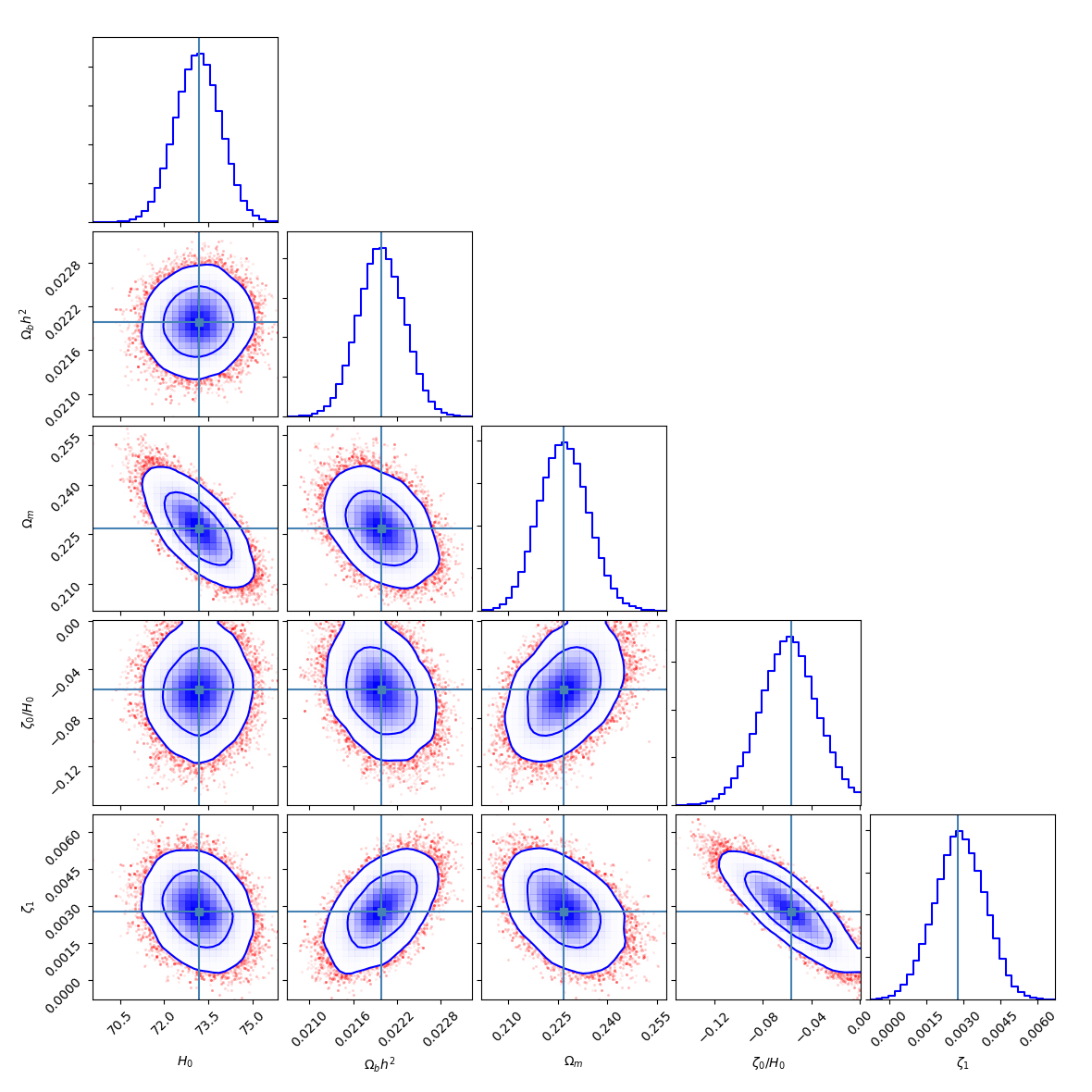}
    \caption{Constraints on the parameters from $1\sigma$ to $2\sigma$ confidence level obtained by the fitting procedure.}
    \label{hist}
\end{figure*}

We perform a cosmological fit for the viscous model in $f(T)$ gravity based on a Markov chain Monte Carlo method
using the datasets mentioned above.
As references, we also perform the same fitting procedure for the concord cosmological model $\Lambda$CDM,
and a viscous model that is described by Eq. \eqref{impmatters}
but still leaves the dark energy to the mysterious cosmological constant (denoted as v$\Lambda$CDM,
for detailed analysis of similar models, see Ref. \cite{Velten:2013qna,Brevik:2017msy,Hu:2020xus}).
The fitting results and best-fitted parameters are listed in Table \ref{bestfit},
where $\Omega_{b0}$ and $\Omega_{m0}$ are the values of $\Omega_b$ and $\Omega_m$ at present time, respectively,
and $h=H_0/(100\:\mathrm{km/s/Mpc})$.
In the fitting procedure, we consider the radiation density parameter at present $\Omega_{r0}=\Omega_{\gamma 0}(1+0.2271N_\text{eff})$,
where the photon density parameter at present is fixed $\Omega_{\gamma 0}=2.469\times 10^{-5}h^{-2}$ in this work
and $N_\text{eff}$ is the effective number of neutrino species with $N_\text{eff}=3.046$ for the current standard value.
Figure \ref{hist} illustrates the $1\sigma(68.3\%)$ and $2\sigma(95.4\%)$ confidence levels of the parameters of our model.

\begin{table*}[htbp]
    \centering\begin{tabular}{lccc}
        \hline
        \hline
        Parameters       & $\Lambda$CDM                 & v$\Lambda$CDM                 & our model                     \\
        \hline
        $\Omega_{b0}h^2$ & $0.0222^{+0.0001}_{-0.0001}$ & $0.0222^{+0.0003}_{-0.0003}$  & $0.0220^{+0.0003}_{-0.0003}$  \\
        $\Omega_{m0}$    & $0.2529^{+0.0019}_{-0.0027}$ & $0.2519^{+0.0079}_{-0.0080}$  & $0.2268^{+0.0073}_{-0.0071}$  \\
        $H_0$ (km/s/Mpc) & $68.39^{+0.20}_{-0.22}$      & $68.29^{+0.72}_{-0.71}$       & $73.15^{+0.74}_{-0.75}$       \\
        $\zeta_0/H_0$    & -                            & $-0.0014^{+0.0013}_{-0.0038}$ & $-0.0195^{+0.0077}_{-0.0076}$ \\
        $\zeta_1$        & -                            & $0.0001^{+0.0008}_{-0.0003}$  & $0.0029^{+0.0010}_{-0.0010}$  \\
        \hline
        \hline
    \end{tabular}
    \caption{Best-fitted parameters for $\Lambda$CDM, v$\Lambda$CDM, and our viscous model in $f(T)$ gravity.}
    \label{bestfit}
\end{table*}

\section{Cosmological discussions}
\label{implication}

\subsection{Matter density parameters}
\begin{figure}[htpb]
    \centering
    \includegraphics[width=\linewidth]{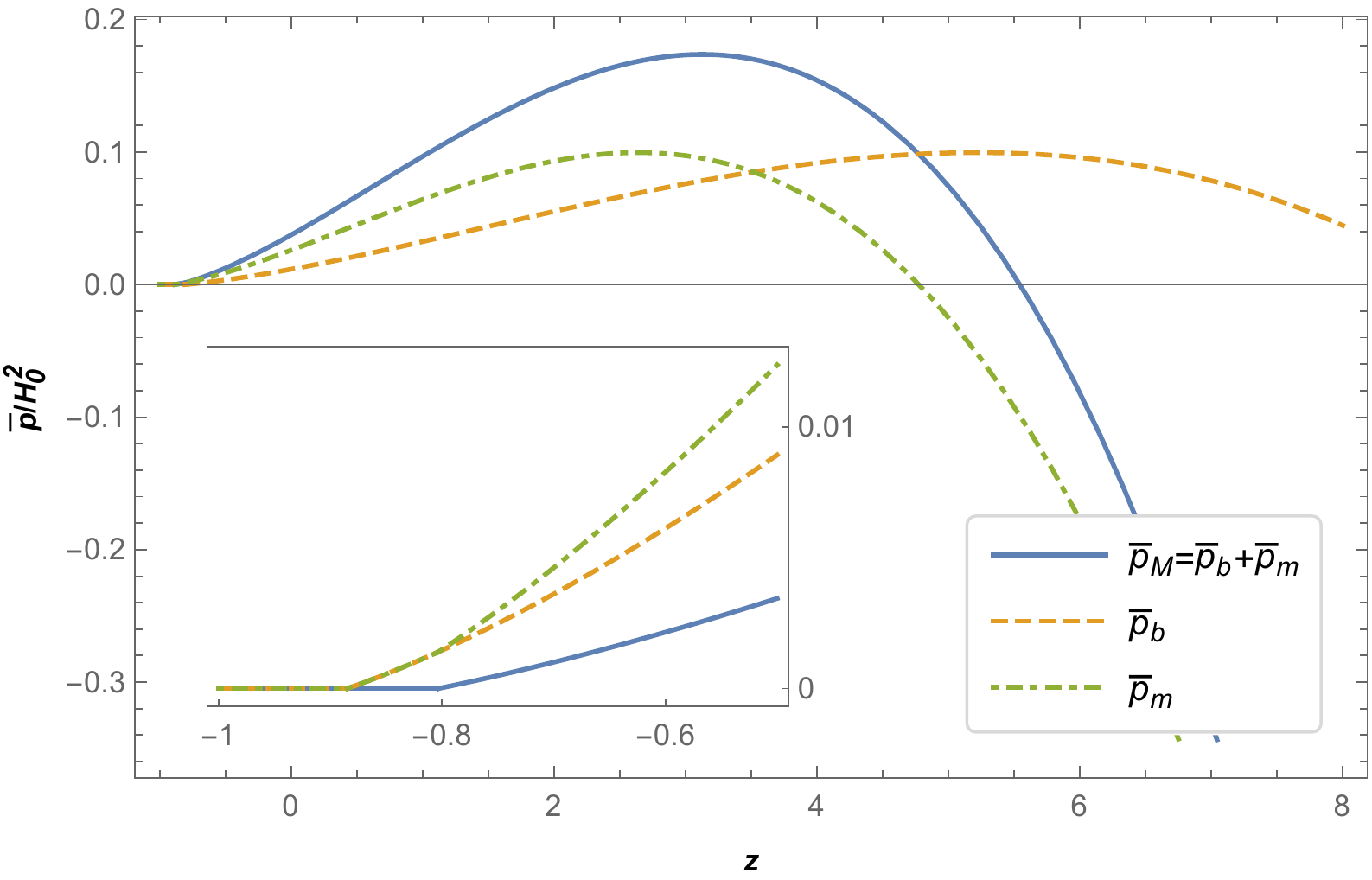}
    \caption{Evolutions of the effective pressures of the total matter, baryonic matter and dark matter,
        plotted in unit of $H_0^2$.}
    \label{p_evo}
\end{figure}

\begin{figure}[htpb]
    \centering
    \includegraphics[width=\linewidth]{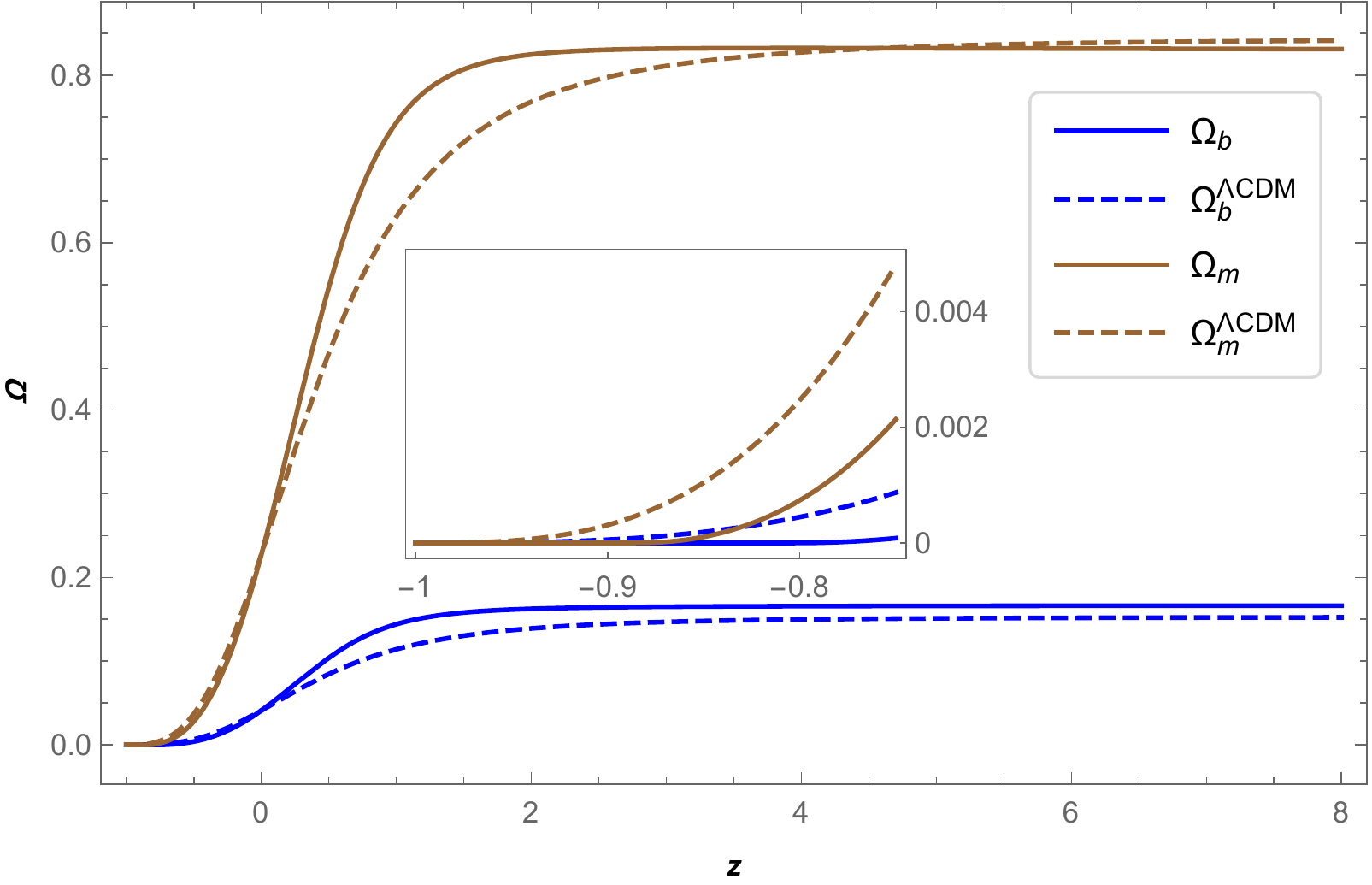}
    \caption{Evolutions of density parameters $\Omega_b$ and $\Omega_m$.}
    \label{bm_evo}
\end{figure}

As shown in Table \ref{bestfit}, the two viscous coefficients in our model, $\zeta_0$ and $\zeta_1$, have different signs,
which, according to Eq. \eqref{impmatters}, means the effective pressures $\bar p$ may have different signs depending on the densities.
And Eq. \eqref{source2} shows that the signs of $\bar p$ indicate the direction of energy transference.
So, the matters may acquire or lose energy due to the universe expansion through viscosity at different cosmic time.
Moreover, the viscous model \eqref{impmatters} suggests that
the baryonic and dark matters will acquire or lose energy at different rates
once they have different densities.
Figures \ref{p_evo} and \ref{bm_evo} illustrate the evolutions of the effective pressures and density parameters of the baryonic and dark matters,
where in the latter the standard $\Lambda$CDM case has also been plotted for reference.


At early time (large redshift $z$), the effective pressures are negative and the matters acquire energy from the expansion,
while at later time (smaller redshift $z$), the pressures are positive and the matters lose energy to the cosmic expansion.
The changes of signs of the effective pressures happen at $z=5.54462,\:8.87689$, and $4.77607$
for total matter, baryonic matter, and dark matter, respectively.
In the future ($-1<z<0$), the effective pressures will return to zero because the matters dissipate and the densities vanish successively,
which will happen at $z=-0.803174$, and $-0.884895$
for the baryonic and dark matters, respectively.

\begin{figure}[htpb]
    \centering
    \includegraphics[width=\linewidth]{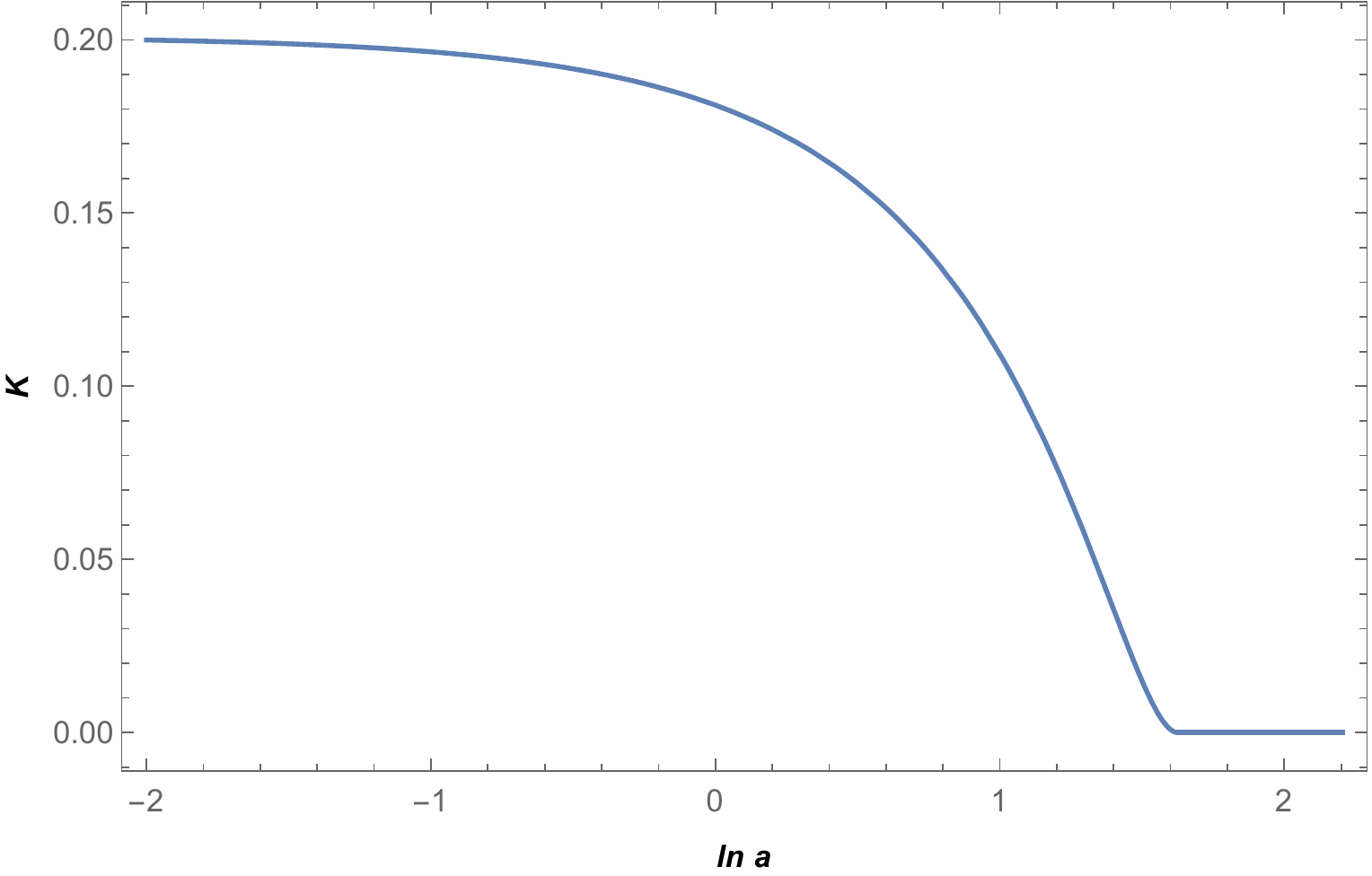}
    \caption{Evolution of the ratio between the baryonic and dark matter density parameters.}
    \label{bm_ratio}
\end{figure}

As mentioned before,
the different paces of evolutions of the baryonic and dark matters will lead to an evolving ratio $K$ between the baryonic and dark matters,
which is plotted in Fig. \ref{bm_ratio}.
In the future, when $\Omega_b$ reaches zero at $z=-0.803174$, $K$ vanishes.
In the remote past, $K=\Omega_b/\Omega_m=0.2009$, which is in fact inferred from the power spectrum of CMB.
On the other hand, while $K$ is difficult to observe,
the baryon-to-photon ratio $\eta$ plays a significant role in the light element abundances.
And since the photon number in the universe can be directly observed,
the visible difference in the evolutions of $\Omega_b$ between our model and the standard $\Lambda$CDM model
may then be subjected to the test of observations.
There are several epochs of the cosmic evolutionary history where $\eta$ can be determined independently.
For the Big Bang nucleosynthesis epoch ($z\sim 10^9$), $\eta$ can be derived from the abundances of primordial elements (see, e.g., \cite{Sarkar:2002er}).
For the recombination epoch ($z\sim1100$), inference of $\eta$ from CMB in fact has already been taken into account in this work when we include the CMB and BAO datasets.
At later time ($2\lesssim z\lesssim 6$),
the abundances of light elements, and hence $\eta$, can be determined from the so-called Lyman $\alpha$ forest
(see, e.g., Refs. \cite{TYTLER2000409,Thuan:2001zc,Coc:2002tr}),
where for the standard model, the cosmic gas doing the absorption is considered nearly primordial
in that it has not been processed during many generations of stars.
In our model, the energy transference between the baryonic matter and the cosmic expansion leads to
the deviation of $\Omega_b$ from the standard model,
which may result in possibly observable difference of such absorption.

Most of these results rely on the modeling of the evolutionary history as a whole
so that one can see the effects of the unified form of viscosities in different eras.
Moreover, the different evolutionary paces of the baryonic and dark matters
essentially come from the separation of viscosities of the two matters in our model,
which makes it possible to discern or even observe the effects of viscosities.
However, the time-dependent ratio between the baryonic and dark matters is not a special result of
simply setting the same form of bulk viscosities for them.
In fact, as long as the baryonic and dark matters do not interact through viscosity
(so that dark matter is still dark and prerecombination BAO can happen),
they most likely feel the viscosity differently,
and acquire/lose energy at different paces.
The model with synchronized evolution and constant ratio is actually the special one and possibly needs some fine tuning.

\subsection{Evolutionary history}
\begin{figure}[htpb]
    \centering
    \includegraphics[width=\linewidth]{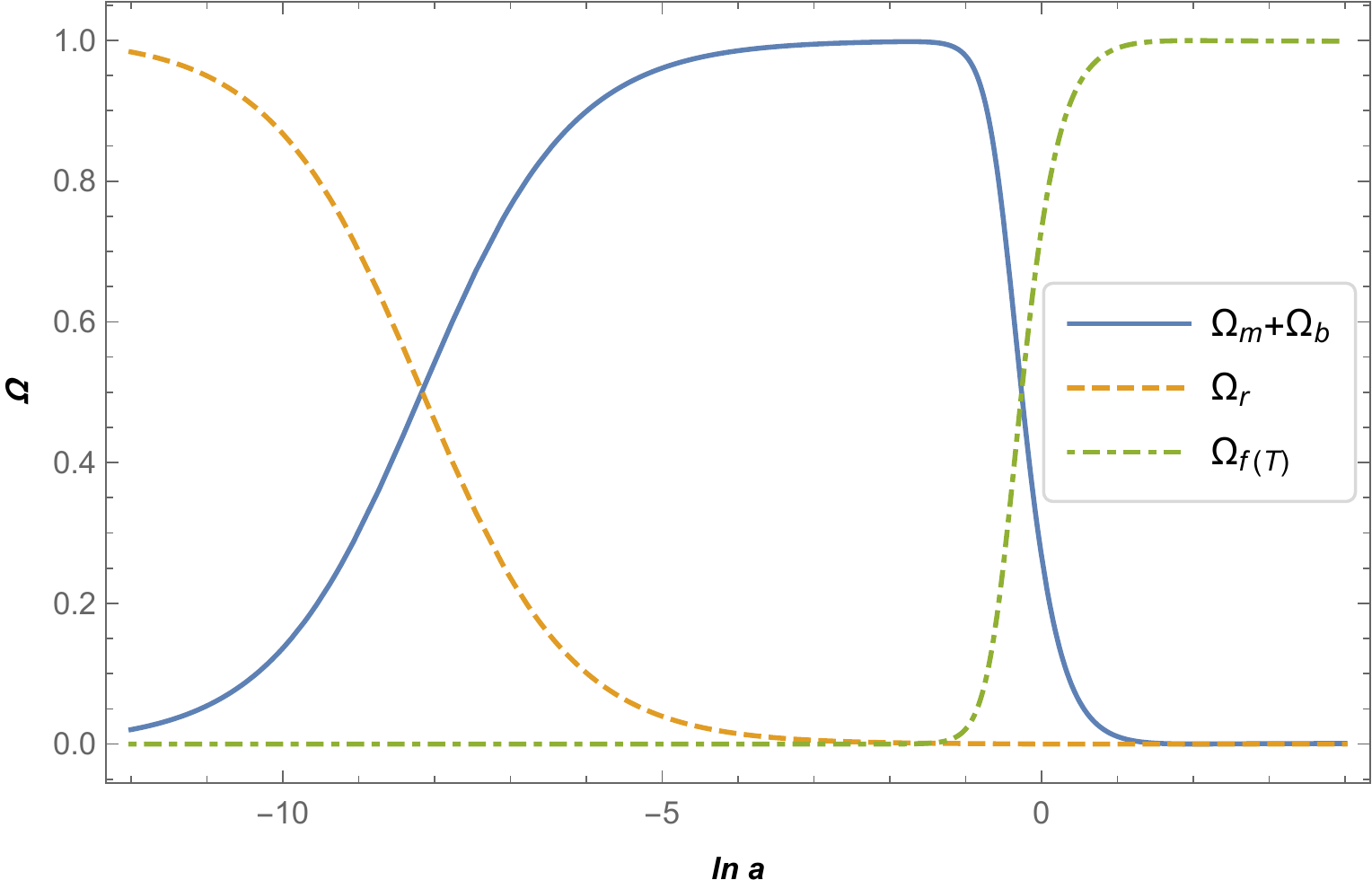}
    \caption{The evolutions of the density parameters.
        $\Omega_{f(T)}$ denotes the effective density parameter of the $f(T)$ modification.}
    \label{den_evo}
\end{figure}

Due to the rather stringent constraints from the CMB dataset \cite{SDSS:2014iwm,Planck:2018vyg},
the best-fitted parameters for $\Lambda$CDM is quite close to the Planck CMB results \cite{Planck:2018vyg}
even though we have included late time datasets of SNIa and Hz.
To check the evolutionary history of the universe,
one can, as usually done in the cosmological study of modified gravities,
consider the extra geometric term from the $f(T)$ modification as an effective fluid.
Then, this $f(T)$ fluid has its own density, namely,
\begin{equation}
    \label{effDE}
    \rho_{f(T)}\equiv \frac{\alpha}{4H^2}.
\end{equation}
Using the best-fitted parameters in Table \ref{bestfit},
we illustrate the evolutions of the density parameters $\Omega = \rho/3H^2$ in Fig. \ref{den_evo}.
The radiation dominating era, matter dominating era, and late time accelerating era are obviously recovered.
Note that in the future,
the density parameters of the baryonic and dark matters will vanish in finite time due to the existence of viscosities,
as discussed in the previous subsection.

\begin{figure}[htpb]
    \centering
    \includegraphics[width=\linewidth]{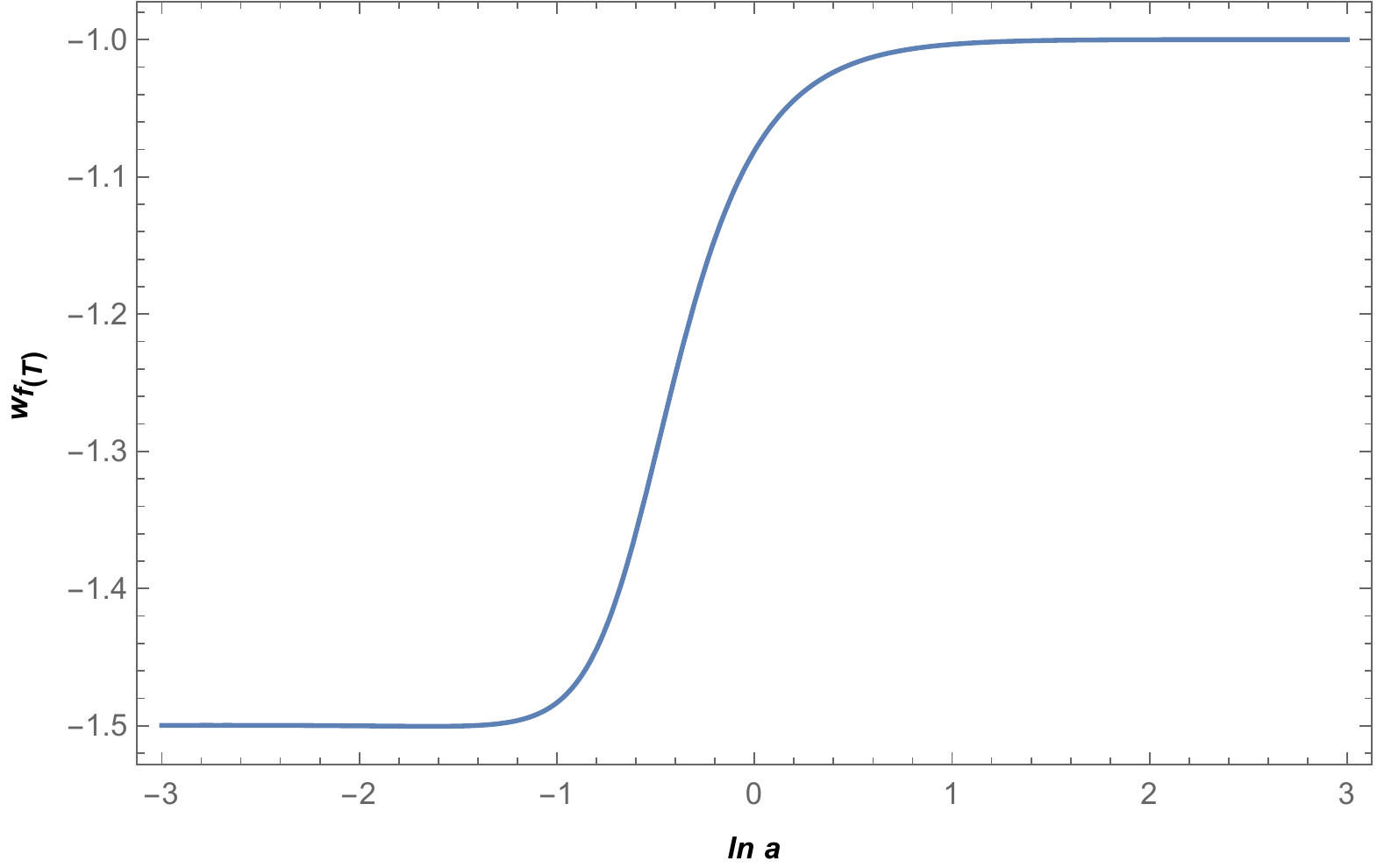}
    \caption{The evolution of the EOS of
        the geometric term from $f(T)$ modification.
        The best-fitted parameters given in Table \ref{bestfit} have be used.}
    \label{wevofig}
\end{figure}

The EOS of the effective $f(T)$ fluid is given by
\begin{equation}
    \label{wde}
    w_{f(T)}=-1-\frac13 \frac{\diff \ln\rho_{f(T)}}{\diff\ln a},
\end{equation}
the evolution of which is illustrated in Fig. \ref{wevofig}.
At present time with $z=0$ and $\ln a=0$, $w_{f(T)}^{(0)}=-1.081$,
whereas the EOS of matter $w_M^{(0)}=0.140$.
Or, equivalently, the density parameter of the $f(T)$ fluid $\Omega_{f(T)}=0.732$,
while the contribution from the effective pressures of the viscosities $\bar p_M/3H_0^2=0.012$.
This suggests that the main contribution to the accelerating expansion at present time comes from the $f(T)$ modification.
In fact, the viscosity model \eqref{impmatters} we have introduced in this work allows for consideration of the physics of the prerecombination era,
for which the observational datasets give rather stringent constraints and suppress the possible magnitude of the viscosities.

At future infinity with $z\rightarrow-1$ and $\ln a\rightarrow+\infty$,
$w_{f(T)}$ approach to $-1$,
indicating a de Sitter fate of the universe.
Before that, the matter densities will reach zero in the finite future.

Therefore, we can conclude that the inclusion of bulk viscosities of the matters
and considering the $f(T)$ modification of gravity as the main drive of the present acceleration
will not violate the known history of the universe.
The density parameters of the baryonic and dark matters will vanish in the finite future due to the viscosities.
The universe will continue to evolve without them,
and will finally enter a de Sitter expansion phase under the effect of $f(T)$ modification.

\subsection{$f(T)$ modification as an effective dark energy}
\begin{figure}[htpb]
    \centering
    \includegraphics[width=\linewidth]{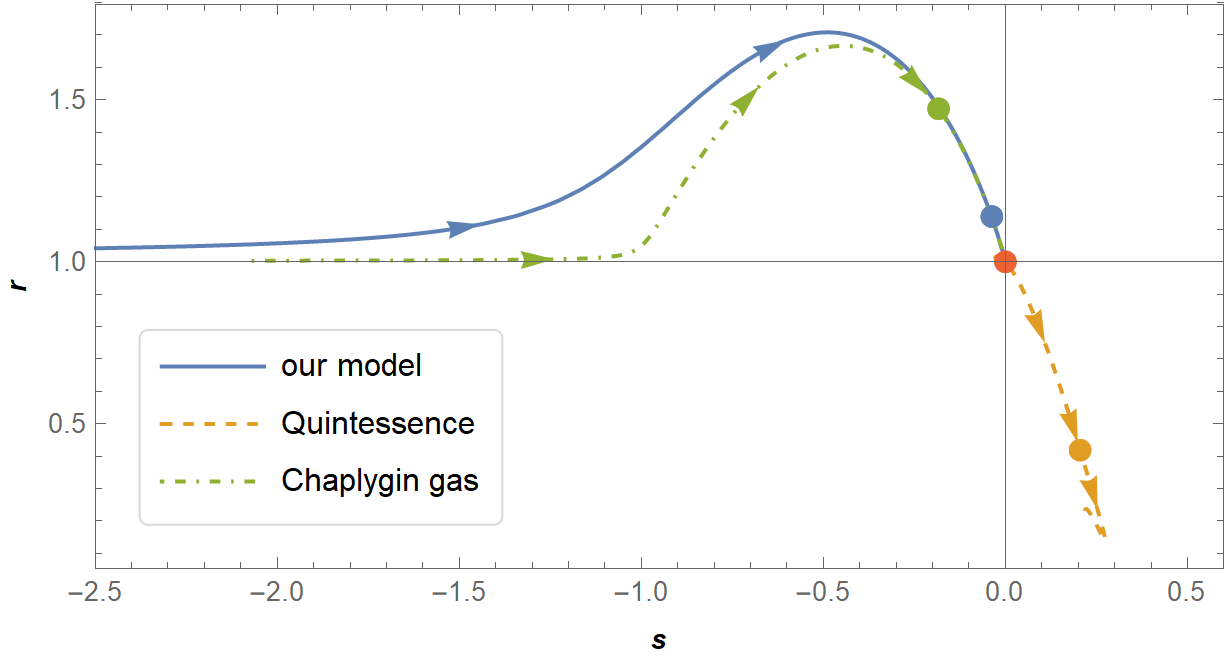}
    \caption{The statefinder pair $(r,s)$ for our model,
        as well as a specific quintessence model and a Chaplygin gas model described in \cite{Sahni:2002fz}.
        The solid rounded points on the lines indicate the present states.
        The solid rounded point at $(0,1)$ denotes the $\Lambda$CDM model.}
    \label{fig:rs}
\end{figure}

Although the $f(T)$ model that drives the accelerated expansion is not considered as a dynamical field in this work,
it effectively plays the role of dark energy and does evolve over the cosmic time.
To discriminate different time-varying models of dark energy or effective dark energy,
the statefinder parameters $(r,s)$ have been introduced \cite{Sahni:2002fz}.
This pair $(r,s)$ is defined as
\begin{equation}
    \label{rs}
    r=\frac{\dddot{a}}{aH^{3}},\quad s=\frac{r-1}{3\left(q-\frac{1}{2}\right)},
\end{equation}
with $q=-{a\ddot{a} }/{\dot{a}^{2} }$ being the deceleration parameter.
Since $H$ and $q$ are composed from the scale factor $a$ and its first and second order of derivatives,
the geometric parameter $r$ is then naturally related to the next order of derivative.
And $s$ is a combination of $r$ and $q$.
Based on the values of the pair $(r,s)$,
the (effective) dark energy models can be put into three categories:
\begin{itemize}
    \item[$\bullet$] $r>1,\:s<0$, Chaplygin gas model;
    \item[$\bullet$] $r<1,\:s>0$, quintessence models;
    \item[$\bullet$] $r=1,\:s=0$, the standard $\Lambda$CDM model.
\end{itemize}
Using the best-fitted parameters in Table \ref{bestfit},
we plot the $r$-$s$ diagram of our model in Fig. \ref{fig:rs},
where we also plot the specific quintessence and Chaplygin gas models described in Ref. \cite{Sahni:2002fz} for comparison.
One can see that our model mimics a kind of Chaplygin gas model,
and eventually it will arrive at the $\Lambda$CDM point $(1,0)$.
Therefore, the late time behavior described in the previous subsection,
i.e., that the universe is currently experiencing a phantom-like dark energy but eventually will enter a de Sitter phase,
is originated from the feature of the $f(T)$ model.

\begin{figure}[htpb]
    \centering
    \includegraphics[width=\linewidth]{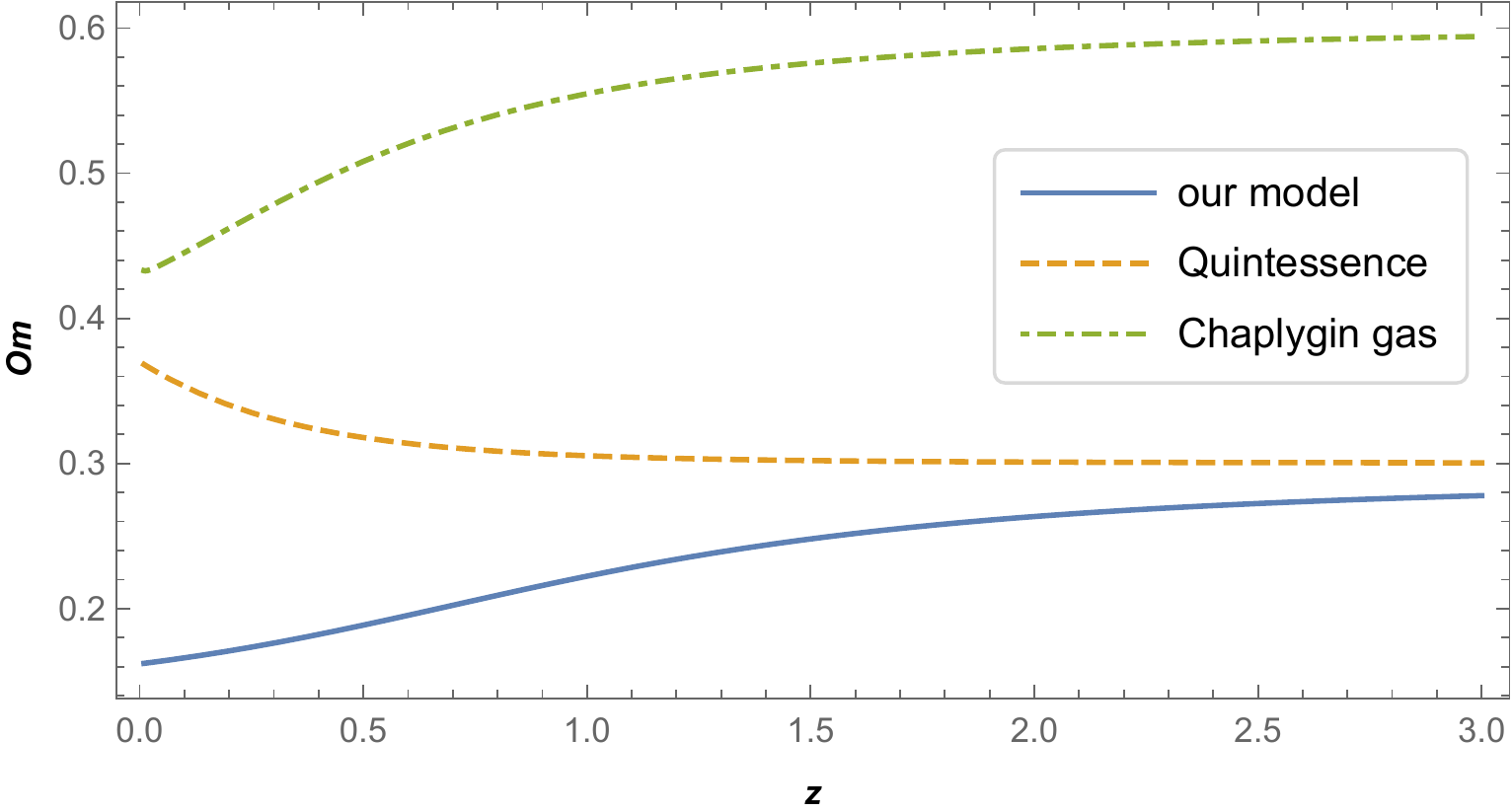}
    \caption{$Om$-$z$ diagram for our model,
        as well as a specific quintessence model and a Chaplygin gas model described in \cite{Sahni:2002fz} for contrast.
    }
    \label{fig:omz}
\end{figure}

Another useful tool to tell apart time-dependent dark energy models is the $Om$ diagnostic \cite{Sahni:2008xx}.
$Om$ is defined as
\begin{equation}
    \label{om}
    Om\left ( z \right ) =\frac{H^{2}\left ( z \right )/H_0^2 -1}{\left ( 1+z \right 	)^{3}-1 },
\end{equation}
which provides a null test of the (effective) dark energy being a cosmological constant $\Lambda$.
The sloping of the trajectory on the $Om$-$z$ diagram is used to categorize the different types of (effective) dark energy:
\begin{itemize}
    \item[$\bullet$] a negative slope indicates a quintessence type of dark energy;
    \item[$\bullet$] a positive slope indicates a phantom-like dark energy;
    \item[$\bullet$] $\Lambda$CDM model gives a horizontal line.
\end{itemize}
In Fig. \ref{fig:omz} we plot the $Om$-$z$ diagram of our model, as well as those of
a specific quintessence model and a Chaplygin gas model described in Ref. \cite{Sahni:2002fz} for comparison.
Obviously, the viscous $f(T)$ model under consideration provides a phantom-like effective dark energy like the Chaplygin gas model,
which is in accordance with the effective EOS $w_{f(T)}$ shown in Fig. \ref{wevofig}.

\subsection{Hubble tension}
Recently, more precise observations seem to reveal that
there are discrepancies among the values of the cosmological parameters fitted by the different datasets
(see, e.g., Ref. \cite{Freedman:2017yms} for a historical review).
One of the most shocking tensions is between the inferences of $H_0$ from the Planck CMB data, reporting $H_0=67.4\pm0.5\:\mathrm{km/(s\:Mpc)}$ \cite{Planck:2018vyg},
and that from local distance ladder measurements by Hubble Space Telescope (HST) which
reports $H_0=74.03\pm1.42\:\mathrm{km/(s\:Mpc)}$ \cite{Riess:2019cxk}.
The two different inferred values of $H_0$ are at a $4.4\sigma$ difference.
From Table \ref{bestfit}, another immediate observation is that
in our model $H_0$ is raised to a value close to the astrophysical measurements by HST.
Comparing with the v$\Lambda$CDM column, one can see that
adding bulk viscosity corrections to the material contents in the $\Lambda$CDM does not help with this tension very much.
A semi-quantitative explanation is that an extra geometric term in the form $1/T\sim 1/H^2$
makes the sloping of $H$ over the cosmic time more gently than a dark content with a constant energy density.
This changes both the sound horizon $r_s^*$ given in \eqref{rs} and the angular diameter distance to the last-scattering surface
\begin{equation}
    \label{da}
    d_A^*=\int_0^{z^*}\frac{\diff z}{H(z)},
\end{equation}
since both of which are integrands of $1/H$ over the cosmic time (represented by the redshift $z$).
While the peak spacing of the CMB spectrum determines quite precisely the ratio $\theta^*=r_s^*/d_A^*$,
a more gently sloping $H$ may then lead to a larger $H_0$.
In this sense, the alleviation of the Hubble tension in our model in fact comes from the $1/T$ term of modification.
In the current work, the new model of viscosity allows one to describe the early and late time evolution as a whole,
which makes it possible to compare the local and global measurements of $H_0$ and hence to discuss Hubble tension problem.
However, the viscosity terms that are not negligible in this model also change the physics before the last scattering a little bit,
giving a slightly smaller sound speed of the photon-baryon fluid and hence a smaller sound horizon $r_s^*$,
which in turn slightly lower the $H_0$ value.
Nonetheless, the fitting result shows that the inclusion of viscosities does not diminish the model's successful alleviation of the Hubble tension.

\section{Concluding remarks}
In this work, we have studied the bulk viscous cosmology under the framework of $f(T)$ gravity.
In order to consider the whole evolutionary history including both the early time prerecombination physics and late time acceleration,
we have proposed a new simple viscous model for the baryonic and dark matters that absorbs the merits of the viscous models in the literature.
Due to the stringent constraints from the observations,
viscosities are not likely significant enough to account for the accelerating expansion of the universe,
we therefore treat the viscosities as the corrections of imperfectness to the fluid models of the baryonic and dark matters,
and leave the major drive of the acceleration to the $f(T)$ modification.

Using the CMB, BAO, Hz, and the Pantheon compilation of SNIa data,
we have performed an observational fit for the cosmological and model parameters.
The fitting result shows that the two viscous coefficients in our model have different signs,
which means both the baryonic and dark matters will lose or acquire energy due to the viscosities at different times.
Since the viscosity model considered in this work is related to the material density,
the baryonic and dark matters will have different rates of energy transference,
and the ratio between the densities of them will not be a constant.
Both the baryonic and dark matters acquire energy to the cosmic expansion at early time,
while they lose energy at late time.
The changes of directions of energy transference happen at redshifts $z=8.87689$, and $4.77607$ for the baryonic and dark matters, respectively.
In the future, they will vanish in finite time due to dissipation, which will happen at $z=-0.803174$ and $-0.884895$
for the baryonic and dark matters, respectively.

With the fitting result, it is shown that our model can recover the known history of the universe
including the radiation dominating era, matter dominating era, and the acceleration era.
The main contribution to the accelerating expansion comes from the $f(T)$ fluid of the modification.
The diagnostic analyses show that our model mimics a phantom-like dark energy.
Nonetheless, at the future infinity,
the universe will eventually enter a de Sitter expansion phase.

The Hubble constant $H_0$ is fitted to be $73.15$ ( km/s/Mpc ),
significantly alleviating the Hubble tension.
This is mainly because the $f(T)$ modification makes the sloping of $H$ over the cosmic time more gently,
which raises the value of $H_0$ while fitting the CMB peak spacings.
The viscosity in fact acts in an opposite direction in this problem,
namely, it makes the sound horizon $r_s^*$ slightly smaller and hence, lowers $H_0$ a little bit.

Conceptually, viscosity arises from the interaction between material particles.
However, a viscosity model that contains only the geometric part, e.g., Eq. \eqref{zetaH},
will lead to a corresponding term in the field equation that is purely geometrical,
which can in effect be interpreted as some modification to the spacetime geometry.
That is, starting from some modified gravity model without viscosity, one may arrive in the same field equation.
Or, one can understand this the other way around.
The viscosity may be yet another physical interpretation of modified gravity.
Similarly, a viscous model involves both the energy of the matter and the spacetime geometry
can also be interpreted as some modified gravity that considers nonminimal interactions between matter and gravity.
For example, a nonminimal coupling $f(T)$ model without viscosity (see, e.g., Refs. \cite{Lin:2016nvj,Lin:2018xzd})
may have the same behavior as the viscous model considered in this work.
This is also referred to as the degeneracy of the viscous cosmologies \cite{Velten:2013qna}.

Nonetheless, viscosity is one of the realistic ways to consider material interactions of the cosmic contents.
The viscous model we employ in this work, although based on the models in the literature,
is chosen for simplicity.
And we have also assumed that it has the same form and has the same coefficients for both the baryonic and dark matters.
A more sophistic model may start from the thermodynamic theory that takes the dark matter model into account,
and should leave the viscous coefficients all to the observational fit.
Moreover, beyond this study of background evolution of the cosmological model with viscosities in $f(T)$ gravity,
the growth of perturbation should also be interesting once viscosity is not negligible \cite{Gagnon:2011id}.
Future works can be done considering these issues.

\section*{Acknowledgement}
\label{ackn}
This work is supported by the National Science Foundation of China under Grant No. 12105179.

\bibliography{ref}

\begin{thebibliography}{78}%
\makeatletter
\providecommand \@ifxundefined [1]{%
 \@ifx{#1\undefined}
}%
\providecommand \@ifnum [1]{%
 \ifnum #1\expandafter \@firstoftwo
 \else \expandafter \@secondoftwo
 \fi
}%
\providecommand \@ifx [1]{%
 \ifx #1\expandafter \@firstoftwo
 \else \expandafter \@secondoftwo
 \fi
}%
\providecommand \natexlab [1]{#1}%
\providecommand \enquote  [1]{``#1''}%
\providecommand \bibnamefont  [1]{#1}%
\providecommand \bibfnamefont [1]{#1}%
\providecommand \citenamefont [1]{#1}%
\providecommand \href@noop [0]{\@secondoftwo}%
\providecommand \href [0]{\begingroup \@sanitize@url \@href}%
\providecommand \@href[1]{\@@startlink{#1}\@@href}%
\providecommand \@@href[1]{\endgroup#1\@@endlink}%
\providecommand \@sanitize@url [0]{\catcode `\\12\catcode `\$12\catcode
  `\&12\catcode `\#12\catcode `\^12\catcode `\_12\catcode `\%12\relax}%
\providecommand \@@startlink[1]{}%
\providecommand \@@endlink[0]{}%
\providecommand \url  [0]{\begingroup\@sanitize@url \@url }%
\providecommand \@url [1]{\endgroup\@href {#1}{\urlprefix }}%
\providecommand \urlprefix  [0]{URL }%
\providecommand \Eprint [0]{\href }%
\providecommand \doibase [0]{https://doi.org/}%
\providecommand \selectlanguage [0]{\@gobble}%
\providecommand \bibinfo  [0]{\@secondoftwo}%
\providecommand \bibfield  [0]{\@secondoftwo}%
\providecommand \translation [1]{[#1]}%
\providecommand \BibitemOpen [0]{}%
\providecommand \bibitemStop [0]{}%
\providecommand \bibitemNoStop [0]{.\EOS\space}%
\providecommand \EOS [0]{\spacefactor3000\relax}%
\providecommand \BibitemShut  [1]{\csname bibitem#1\endcsname}%
\let\auto@bib@innerbib\@empty
\bibitem [{\citenamefont {Copeland}\ \emph {et~al.}(2006)\citenamefont
  {Copeland}, \citenamefont {Sami},\ and\ \citenamefont
  {Tsujikawa}}]{Copeland:2006wr}%
  \BibitemOpen
  \bibfield  {author} {\bibinfo {author} {\bibfnamefont {E.~J.}\ \bibnamefont
  {Copeland}}, \bibinfo {author} {\bibfnamefont {M.}~\bibnamefont {Sami}},\
  and\ \bibinfo {author} {\bibfnamefont {S.}~\bibnamefont {Tsujikawa}},\
  }\bibfield  {title} {\bibinfo {title} {{Dynamics of dark energy}},\ }\href
  {https://doi.org/10.1142/S021827180600942X} {\bibfield  {journal} {\bibinfo
  {journal} {Int. J. Mod. Phys. D}\ }\textbf {\bibinfo {volume} {15}},\
  \bibinfo {pages} {1753} (\bibinfo {year} {2006})},\ \Eprint
  {https://arxiv.org/abs/hep-th/0603057} {arXiv:hep-th/0603057} \BibitemShut
  {NoStop}%
\bibitem [{\citenamefont {Bahamonde}\ \emph {et~al.}(2018)\citenamefont
  {Bahamonde}, \citenamefont {B\"ohmer}, \citenamefont {Carloni}, \citenamefont
  {Copeland}, \citenamefont {Fang},\ and\ \citenamefont
  {Tamanini}}]{Bahamonde:2017ize}%
  \BibitemOpen
  \bibfield  {author} {\bibinfo {author} {\bibfnamefont {S.}~\bibnamefont
  {Bahamonde}}, \bibinfo {author} {\bibfnamefont {C.~G.}\ \bibnamefont
  {B\"ohmer}}, \bibinfo {author} {\bibfnamefont {S.}~\bibnamefont {Carloni}},
  \bibinfo {author} {\bibfnamefont {E.~J.}\ \bibnamefont {Copeland}}, \bibinfo
  {author} {\bibfnamefont {W.}~\bibnamefont {Fang}},\ and\ \bibinfo {author}
  {\bibfnamefont {N.}~\bibnamefont {Tamanini}},\ }\bibfield  {title} {\bibinfo
  {title} {{Dynamical systems applied to cosmology: dark energy and modified
  gravity}},\ }\href {https://doi.org/10.1016/j.physrep.2018.09.001} {\bibfield
   {journal} {\bibinfo  {journal} {Phys. Rept.}\ }\textbf {\bibinfo {volume}
  {775-777}},\ \bibinfo {pages} {1} (\bibinfo {year} {2018})},\ \Eprint
  {https://arxiv.org/abs/1712.03107} {arXiv:1712.03107 [gr-qc]} \BibitemShut
  {NoStop}%
\bibitem [{\citenamefont {Frusciante}\ and\ \citenamefont
  {Perenon}(2020)}]{Frusciante:2019xia}%
  \BibitemOpen
  \bibfield  {author} {\bibinfo {author} {\bibfnamefont {N.}~\bibnamefont
  {Frusciante}}\ and\ \bibinfo {author} {\bibfnamefont {L.}~\bibnamefont
  {Perenon}},\ }\bibfield  {title} {\bibinfo {title} {{Effective field theory
  of dark energy: A review}},\ }\href
  {https://doi.org/10.1016/j.physrep.2020.02.004} {\bibfield  {journal}
  {\bibinfo  {journal} {Phys. Rept.}\ }\textbf {\bibinfo {volume} {857}},\
  \bibinfo {pages} {1} (\bibinfo {year} {2020})},\ \Eprint
  {https://arxiv.org/abs/1907.03150} {arXiv:1907.03150 [astro-ph.CO]}
  \BibitemShut {NoStop}%
\bibitem [{\citenamefont {Ishak}(2019)}]{Ishak:2018his}%
  \BibitemOpen
  \bibfield  {author} {\bibinfo {author} {\bibfnamefont {M.}~\bibnamefont
  {Ishak}},\ }\bibfield  {title} {\bibinfo {title} {{Testing General Relativity
  in Cosmology}},\ }\href {https://doi.org/10.1007/s41114-018-0017-4}
  {\bibfield  {journal} {\bibinfo  {journal} {Living Rev. Rel.}\ }\textbf
  {\bibinfo {volume} {22}},\ \bibinfo {pages} {1} (\bibinfo {year} {2019})},\
  \Eprint {https://arxiv.org/abs/1806.10122} {arXiv:1806.10122 [astro-ph.CO]}
  \BibitemShut {NoStop}%
\bibitem [{\citenamefont {De~Felice}\ and\ \citenamefont
  {Tsujikawa}(2010)}]{DeFelice:2010aj}%
  \BibitemOpen
  \bibfield  {author} {\bibinfo {author} {\bibfnamefont {A.}~\bibnamefont
  {De~Felice}}\ and\ \bibinfo {author} {\bibfnamefont {S.}~\bibnamefont
  {Tsujikawa}},\ }\bibfield  {title} {\bibinfo {title} {{f(R) theories}},\
  }\href {https://doi.org/10.12942/lrr-2010-3} {\bibfield  {journal} {\bibinfo
  {journal} {Living Rev. Rel.}\ }\textbf {\bibinfo {volume} {13}},\ \bibinfo
  {pages} {3} (\bibinfo {year} {2010})},\ \Eprint
  {https://arxiv.org/abs/1002.4928} {arXiv:1002.4928 [gr-qc]} \BibitemShut
  {NoStop}%
\bibitem [{\citenamefont {Sotiriou}\ and\ \citenamefont
  {Faraoni}(2010)}]{Sotiriou:2008rp}%
  \BibitemOpen
  \bibfield  {author} {\bibinfo {author} {\bibfnamefont {T.~P.}\ \bibnamefont
  {Sotiriou}}\ and\ \bibinfo {author} {\bibfnamefont {V.}~\bibnamefont
  {Faraoni}},\ }\bibfield  {title} {\bibinfo {title} {{f(R) Theories Of
  Gravity}},\ }\href {https://doi.org/10.1103/RevModPhys.82.451} {\bibfield
  {journal} {\bibinfo  {journal} {Rev. Mod. Phys.}\ }\textbf {\bibinfo {volume}
  {82}},\ \bibinfo {pages} {451} (\bibinfo {year} {2010})},\ \Eprint
  {https://arxiv.org/abs/0805.1726} {arXiv:0805.1726 [gr-qc]} \BibitemShut
  {NoStop}%
\bibitem [{\citenamefont {Capozziello}\ and\ \citenamefont
  {De~Laurentis}(2011)}]{Capozziello:2011et}%
  \BibitemOpen
  \bibfield  {author} {\bibinfo {author} {\bibfnamefont {S.}~\bibnamefont
  {Capozziello}}\ and\ \bibinfo {author} {\bibfnamefont {M.}~\bibnamefont
  {De~Laurentis}},\ }\bibfield  {title} {\bibinfo {title} {{Extended Theories
  of Gravity}},\ }\href {https://doi.org/10.1016/j.physrep.2011.09.003}
  {\bibfield  {journal} {\bibinfo  {journal} {Phys. Rept.}\ }\textbf {\bibinfo
  {volume} {509}},\ \bibinfo {pages} {167} (\bibinfo {year} {2011})},\ \Eprint
  {https://arxiv.org/abs/1108.6266} {arXiv:1108.6266 [gr-qc]} \BibitemShut
  {NoStop}%
\bibitem [{\citenamefont {Nojiri}\ and\ \citenamefont
  {Odintsov}(2011)}]{Nojiri:2010wj}%
  \BibitemOpen
  \bibfield  {author} {\bibinfo {author} {\bibfnamefont {S.}~\bibnamefont
  {Nojiri}}\ and\ \bibinfo {author} {\bibfnamefont {S.~D.}\ \bibnamefont
  {Odintsov}},\ }\bibfield  {title} {\bibinfo {title} {{Unified cosmic history
  in modified gravity: from F(R) theory to Lorentz non-invariant models}},\
  }\href {https://doi.org/10.1016/j.physrep.2011.04.001} {\bibfield  {journal}
  {\bibinfo  {journal} {Phys. Rept.}\ }\textbf {\bibinfo {volume} {505}},\
  \bibinfo {pages} {59} (\bibinfo {year} {2011})},\ \Eprint
  {https://arxiv.org/abs/1011.0544} {arXiv:1011.0544 [gr-qc]} \BibitemShut
  {NoStop}%
\bibitem [{\citenamefont {Aldrovandi}\ and\ \citenamefont
  {Pereira}(2013)}]{Aldrovandi:2013wha}%
  \BibitemOpen
  \bibfield  {author} {\bibinfo {author} {\bibfnamefont {R.}~\bibnamefont
  {Aldrovandi}}\ and\ \bibinfo {author} {\bibfnamefont {J.~G.}\ \bibnamefont
  {Pereira}},\ }\href {https://doi.org/10.1007/978-94-007-5143-9} {\emph
  {\bibinfo {title} {{Teleparallel Gravity}: {An Introduction}}}},\ Vol.\
  \bibinfo {volume} {173}\ (\bibinfo  {publisher} {Springer},\ \bibinfo {year}
  {2013})\BibitemShut {NoStop}%
\bibitem [{\citenamefont {Maluf}(2013)}]{Maluf2013}%
  \BibitemOpen
  \bibfield  {author} {\bibinfo {author} {\bibfnamefont {J.~W.}\ \bibnamefont
  {Maluf}},\ }\bibfield  {title} {\bibinfo {title} {The teleparallel equivalent
  of general relativity},\ }\href
  {https://doi.org/https://doi.org/10.1002/andp.201200272} {\bibfield
  {journal} {\bibinfo  {journal} {Annalen der Physik}\ }\textbf {\bibinfo
  {volume} {525}},\ \bibinfo {pages} {339} (\bibinfo {year} {2013})},\ \Eprint
  {https://arxiv.org/abs/https://onlinelibrary.wiley.com/doi/pdf/10.1002/andp.201200272}
  {https://onlinelibrary.wiley.com/doi/pdf/10.1002/andp.201200272} \BibitemShut
  {NoStop}%
\bibitem [{\citenamefont {Bengochea}\ and\ \citenamefont
  {Ferraro}(2009)}]{PhysRevD.79.124019}%
  \BibitemOpen
  \bibfield  {author} {\bibinfo {author} {\bibfnamefont {G.~R.}\ \bibnamefont
  {Bengochea}}\ and\ \bibinfo {author} {\bibfnamefont {R.}~\bibnamefont
  {Ferraro}},\ }\bibfield  {title} {\bibinfo {title} {Dark torsion as the
  cosmic speed-up},\ }\href {https://doi.org/10.1103/PhysRevD.79.124019}
  {\bibfield  {journal} {\bibinfo  {journal} {Phys. Rev. D}\ }\textbf {\bibinfo
  {volume} {79}},\ \bibinfo {pages} {124019} (\bibinfo {year}
  {2009})}\BibitemShut {NoStop}%
\bibitem [{\citenamefont {Linder}(2010)}]{PhysRevD.81.127301}%
  \BibitemOpen
  \bibfield  {author} {\bibinfo {author} {\bibfnamefont {E.~V.}\ \bibnamefont
  {Linder}},\ }\bibfield  {title} {\bibinfo {title} {Einstein's other gravity
  and the acceleration of the universe},\ }\href
  {https://doi.org/10.1103/PhysRevD.81.127301} {\bibfield  {journal} {\bibinfo
  {journal} {Phys. Rev. D}\ }\textbf {\bibinfo {volume} {81}},\ \bibinfo
  {pages} {127301} (\bibinfo {year} {2010})}\BibitemShut {NoStop}%
\bibitem [{\citenamefont {Cai}\ \emph {et~al.}(2016)\citenamefont {Cai},
  \citenamefont {Capozziello}, \citenamefont {De~Laurentis},\ and\
  \citenamefont {Saridakis}}]{Cai:2015emx}%
  \BibitemOpen
  \bibfield  {author} {\bibinfo {author} {\bibfnamefont {Y.-F.}\ \bibnamefont
  {Cai}}, \bibinfo {author} {\bibfnamefont {S.}~\bibnamefont {Capozziello}},
  \bibinfo {author} {\bibfnamefont {M.}~\bibnamefont {De~Laurentis}},\ and\
  \bibinfo {author} {\bibfnamefont {E.~N.}\ \bibnamefont {Saridakis}},\
  }\bibfield  {title} {\bibinfo {title} {{f(T) teleparallel gravity and
  cosmology}},\ }\href {https://doi.org/10.1088/0034-4885/79/10/106901}
  {\bibfield  {journal} {\bibinfo  {journal} {Rept. Prog. Phys.}\ }\textbf
  {\bibinfo {volume} {79}},\ \bibinfo {pages} {106901} (\bibinfo {year}
  {2016})},\ \Eprint {https://arxiv.org/abs/1511.07586} {arXiv:1511.07586
  [gr-qc]} \BibitemShut {NoStop}%
\bibitem [{\citenamefont {Nojiri}\ \emph {et~al.}(2017)\citenamefont {Nojiri},
  \citenamefont {Odintsov},\ and\ \citenamefont {Oikonomou}}]{Nojiri:2017ncd}%
  \BibitemOpen
  \bibfield  {author} {\bibinfo {author} {\bibfnamefont {S.}~\bibnamefont
  {Nojiri}}, \bibinfo {author} {\bibfnamefont {S.}~\bibnamefont {Odintsov}},\
  and\ \bibinfo {author} {\bibfnamefont {V.}~\bibnamefont {Oikonomou}},\
  }\bibfield  {title} {\bibinfo {title} {{Modified Gravity Theories on a
  Nutshell: Inflation, Bounce and Late-time Evolution}},\ }\href
  {https://doi.org/10.1016/j.physrep.2017.06.001} {\bibfield  {journal}
  {\bibinfo  {journal} {Phys. Rept.}\ }\textbf {\bibinfo {volume} {692}},\
  \bibinfo {pages} {1} (\bibinfo {year} {2017})},\ \Eprint
  {https://arxiv.org/abs/1705.11098} {arXiv:1705.11098 [gr-qc]} \BibitemShut
  {NoStop}%
\bibitem [{\citenamefont {Eckart}(1940)}]{Eckart:1940te}%
  \BibitemOpen
  \bibfield  {author} {\bibinfo {author} {\bibfnamefont {C.}~\bibnamefont
  {Eckart}},\ }\bibfield  {title} {\bibinfo {title} {{The Thermodynamics of
  irreversible processes. 3.. Relativistic theory of the simple fluid}},\
  }\href {https://doi.org/10.1103/PhysRev.58.919} {\bibfield  {journal}
  {\bibinfo  {journal} {Phys. Rev.}\ }\textbf {\bibinfo {volume} {58}},\
  \bibinfo {pages} {919} (\bibinfo {year} {1940})}\BibitemShut {NoStop}%
\bibitem [{\citenamefont {Israel}(1976)}]{Israel:1976tn}%
  \BibitemOpen
  \bibfield  {author} {\bibinfo {author} {\bibfnamefont {W.}~\bibnamefont
  {Israel}},\ }\bibfield  {title} {\bibinfo {title} {{Nonstationary
  irreversible thermodynamics: A Causal relativistic theory}},\ }\href
  {https://doi.org/10.1016/0003-4916(76)90064-6} {\bibfield  {journal}
  {\bibinfo  {journal} {Annals Phys.}\ }\textbf {\bibinfo {volume} {100}},\
  \bibinfo {pages} {310} (\bibinfo {year} {1976})}\BibitemShut {NoStop}%
\bibitem [{\citenamefont {Hiscock}\ and\ \citenamefont
  {Salmonson}(1991)}]{Hiscock:1991sp}%
  \BibitemOpen
  \bibfield  {author} {\bibinfo {author} {\bibfnamefont {W.~A.}\ \bibnamefont
  {Hiscock}}\ and\ \bibinfo {author} {\bibfnamefont {J.}~\bibnamefont
  {Salmonson}},\ }\bibfield  {title} {\bibinfo {title} {{Dissipative
  Boltzmann-Robertson-Walker cosmologies}},\ }\href
  {https://doi.org/10.1103/PhysRevD.43.3249} {\bibfield  {journal} {\bibinfo
  {journal} {Phys. Rev. D}\ }\textbf {\bibinfo {volume} {43}},\ \bibinfo
  {pages} {3249} (\bibinfo {year} {1991})}\BibitemShut {NoStop}%
\bibitem [{\citenamefont {Ren}\ and\ \citenamefont
  {Meng}(2006{\natexlab{a}})}]{Ren:2006en}%
  \BibitemOpen
  \bibfield  {author} {\bibinfo {author} {\bibfnamefont {J.}~\bibnamefont
  {Ren}}\ and\ \bibinfo {author} {\bibfnamefont {X.-H.}\ \bibnamefont {Meng}},\
  }\bibfield  {title} {\bibinfo {title} {{Modified equation of state, scalar
  field and bulk viscosity in friedmann universe}},\ }\href
  {https://doi.org/10.1016/j.physletb.2006.03.029} {\bibfield  {journal}
  {\bibinfo  {journal} {Phys. Lett. B}\ }\textbf {\bibinfo {volume} {636}},\
  \bibinfo {pages} {5} (\bibinfo {year} {2006}{\natexlab{a}})},\ \Eprint
  {https://arxiv.org/abs/astro-ph/0602462} {arXiv:astro-ph/0602462}
  \BibitemShut {NoStop}%
\bibitem [{\citenamefont {Meng}\ \emph {et~al.}(2007)\citenamefont {Meng},
  \citenamefont {Ren},\ and\ \citenamefont {Hu}}]{Meng:2005jy}%
  \BibitemOpen
  \bibfield  {author} {\bibinfo {author} {\bibfnamefont {X.-H.}\ \bibnamefont
  {Meng}}, \bibinfo {author} {\bibfnamefont {J.}~\bibnamefont {Ren}},\ and\
  \bibinfo {author} {\bibfnamefont {M.-G.}\ \bibnamefont {Hu}},\ }\bibfield
  {title} {\bibinfo {title} {{Friedmann cosmology with a generalized equation
  of state and bulk viscosity}},\ }\href
  {https://doi.org/10.1088/0253-6102/47/2/036} {\bibfield  {journal} {\bibinfo
  {journal} {Commun. Theor. Phys.}\ }\textbf {\bibinfo {volume} {47}},\
  \bibinfo {pages} {379} (\bibinfo {year} {2007})},\ \Eprint
  {https://arxiv.org/abs/astro-ph/0509250} {arXiv:astro-ph/0509250}
  \BibitemShut {NoStop}%
\bibitem [{\citenamefont {Cataldo}\ \emph {et~al.}(2005)\citenamefont
  {Cataldo}, \citenamefont {Cruz},\ and\ \citenamefont
  {Lepe}}]{Cataldo:2005qh}%
  \BibitemOpen
  \bibfield  {author} {\bibinfo {author} {\bibfnamefont {M.}~\bibnamefont
  {Cataldo}}, \bibinfo {author} {\bibfnamefont {N.}~\bibnamefont {Cruz}},\ and\
  \bibinfo {author} {\bibfnamefont {S.}~\bibnamefont {Lepe}},\ }\bibfield
  {title} {\bibinfo {title} {{Viscous dark energy and phantom evolution}},\
  }\href {https://doi.org/10.1016/j.physletb.2005.05.029} {\bibfield  {journal}
  {\bibinfo  {journal} {Phys. Lett. B}\ }\textbf {\bibinfo {volume} {619}},\
  \bibinfo {pages} {5} (\bibinfo {year} {2005})},\ \Eprint
  {https://arxiv.org/abs/hep-th/0506153} {arXiv:hep-th/0506153} \BibitemShut
  {NoStop}%
\bibitem [{\citenamefont {Brevik}\ and\ \citenamefont
  {Gorbunova}(2005)}]{Brevik:2005bj}%
  \BibitemOpen
  \bibfield  {author} {\bibinfo {author} {\bibfnamefont {I.~H.}\ \bibnamefont
  {Brevik}}\ and\ \bibinfo {author} {\bibfnamefont {O.}~\bibnamefont
  {Gorbunova}},\ }\bibfield  {title} {\bibinfo {title} {{Dark energy and
  viscous cosmology}},\ }\href {https://doi.org/10.1007/s10714-005-0178-9}
  {\bibfield  {journal} {\bibinfo  {journal} {Gen. Rel. Grav.}\ }\textbf
  {\bibinfo {volume} {37}},\ \bibinfo {pages} {2039} (\bibinfo {year}
  {2005})},\ \Eprint {https://arxiv.org/abs/gr-qc/0504001}
  {arXiv:gr-qc/0504001} \BibitemShut {NoStop}%
\bibitem [{\citenamefont {Wilson}\ \emph {et~al.}(2007)\citenamefont {Wilson},
  \citenamefont {Mathews},\ and\ \citenamefont {Fuller}}]{PhysRevD.75.043521}%
  \BibitemOpen
  \bibfield  {author} {\bibinfo {author} {\bibfnamefont {J.~R.}\ \bibnamefont
  {Wilson}}, \bibinfo {author} {\bibfnamefont {G.~J.}\ \bibnamefont
  {Mathews}},\ and\ \bibinfo {author} {\bibfnamefont {G.~M.}\ \bibnamefont
  {Fuller}},\ }\bibfield  {title} {\bibinfo {title} {Bulk viscosity, decaying
  dark matter, and the cosmic acceleration},\ }\href
  {https://doi.org/10.1103/PhysRevD.75.043521} {\bibfield  {journal} {\bibinfo
  {journal} {Phys. Rev. D}\ }\textbf {\bibinfo {volume} {75}},\ \bibinfo
  {pages} {043521} (\bibinfo {year} {2007})}\BibitemShut {NoStop}%
\bibitem [{\citenamefont {Colistete}\ \emph {et~al.}(2007)\citenamefont
  {Colistete}, \citenamefont {Fabris}, \citenamefont {Tossa},\ and\
  \citenamefont {Zimdahl}}]{PhysRevD.76.103516}%
  \BibitemOpen
  \bibfield  {author} {\bibinfo {author} {\bibfnamefont {R.}~\bibnamefont
  {Colistete}}, \bibinfo {author} {\bibfnamefont {J.~C.}\ \bibnamefont
  {Fabris}}, \bibinfo {author} {\bibfnamefont {J.}~\bibnamefont {Tossa}},\ and\
  \bibinfo {author} {\bibfnamefont {W.}~\bibnamefont {Zimdahl}},\ }\bibfield
  {title} {\bibinfo {title} {Bulk viscous cosmology},\ }\href
  {https://doi.org/10.1103/PhysRevD.76.103516} {\bibfield  {journal} {\bibinfo
  {journal} {Phys. Rev. D}\ }\textbf {\bibinfo {volume} {76}},\ \bibinfo
  {pages} {103516} (\bibinfo {year} {2007})}\BibitemShut {NoStop}%
\bibitem [{\citenamefont {Li}\ and\ \citenamefont
  {Barrow}(2009)}]{PhysRevD.79.103521}%
  \BibitemOpen
  \bibfield  {author} {\bibinfo {author} {\bibfnamefont {B.}~\bibnamefont
  {Li}}\ and\ \bibinfo {author} {\bibfnamefont {J.~D.}\ \bibnamefont
  {Barrow}},\ }\bibfield  {title} {\bibinfo {title} {Does bulk viscosity create
  a viable unified dark matter model?},\ }\href
  {https://doi.org/10.1103/PhysRevD.79.103521} {\bibfield  {journal} {\bibinfo
  {journal} {Phys. Rev. D}\ }\textbf {\bibinfo {volume} {79}},\ \bibinfo
  {pages} {103521} (\bibinfo {year} {2009})}\BibitemShut {NoStop}%
\bibitem [{\citenamefont {Avelino}\ and\ \citenamefont
  {Nucamendi}(2010)}]{Avelino:2010pb}%
  \BibitemOpen
  \bibfield  {author} {\bibinfo {author} {\bibfnamefont {A.}~\bibnamefont
  {Avelino}}\ and\ \bibinfo {author} {\bibfnamefont {U.}~\bibnamefont
  {Nucamendi}},\ }\bibfield  {title} {\bibinfo {title} {{Exploring a
  matter-dominated model with bulk viscosity to drive the accelerated expansion
  of the Universe}},\ }\href {https://doi.org/10.1088/1475-7516/2010/08/009}
  {\bibfield  {journal} {\bibinfo  {journal} {JCAP}\ }\textbf {\bibinfo
  {volume} {08}},\ \bibinfo {pages} {009}},\ \Eprint
  {https://arxiv.org/abs/1002.3605} {arXiv:1002.3605 [gr-qc]} \BibitemShut
  {NoStop}%
\bibitem [{\citenamefont {Brevik}\ \emph {et~al.}(2011)\citenamefont {Brevik},
  \citenamefont {Elizalde}, \citenamefont {Nojiri},\ and\ \citenamefont
  {Odintsov}}]{PhysRevD.84.103508}%
  \BibitemOpen
  \bibfield  {author} {\bibinfo {author} {\bibfnamefont {I.}~\bibnamefont
  {Brevik}}, \bibinfo {author} {\bibfnamefont {E.}~\bibnamefont {Elizalde}},
  \bibinfo {author} {\bibfnamefont {S.}~\bibnamefont {Nojiri}},\ and\ \bibinfo
  {author} {\bibfnamefont {S.~D.}\ \bibnamefont {Odintsov}},\ }\bibfield
  {title} {\bibinfo {title} {Viscous little rip cosmology},\ }\href
  {https://doi.org/10.1103/PhysRevD.84.103508} {\bibfield  {journal} {\bibinfo
  {journal} {Phys. Rev. D}\ }\textbf {\bibinfo {volume} {84}},\ \bibinfo
  {pages} {103508} (\bibinfo {year} {2011})}\BibitemShut {NoStop}%
\bibitem [{\citenamefont {Gagnon}\ and\ \citenamefont
  {Lesgourgues}(2011)}]{Gagnon:2011id}%
  \BibitemOpen
  \bibfield  {author} {\bibinfo {author} {\bibfnamefont {J.-S.}\ \bibnamefont
  {Gagnon}}\ and\ \bibinfo {author} {\bibfnamefont {J.}~\bibnamefont
  {Lesgourgues}},\ }\bibfield  {title} {\bibinfo {title} {{Dark goo: Bulk
  viscosity as an alternative to dark energy}},\ }\href
  {https://doi.org/10.1088/1475-7516/2011/09/026} {\bibfield  {journal}
  {\bibinfo  {journal} {JCAP}\ }\textbf {\bibinfo {volume} {09}},\ \bibinfo
  {pages} {026}},\ \Eprint {https://arxiv.org/abs/1107.1503} {arXiv:1107.1503
  [astro-ph.CO]} \BibitemShut {NoStop}%
\bibitem [{\citenamefont {Fabris}\ \emph {et~al.}(2011)\citenamefont {Fabris},
  \citenamefont {de~Oliveira},\ and\ \citenamefont {Velten}}]{Fabris:2011wk}%
  \BibitemOpen
  \bibfield  {author} {\bibinfo {author} {\bibfnamefont {J.~C.}\ \bibnamefont
  {Fabris}}, \bibinfo {author} {\bibfnamefont {P.~L.~C.}\ \bibnamefont
  {de~Oliveira}},\ and\ \bibinfo {author} {\bibfnamefont {H.}~\bibnamefont
  {Velten}},\ }\bibfield  {title} {\bibinfo {title} {{Contraints on unified
  models for dark matter and dark energy using H(z)}},\ }\href
  {https://doi.org/10.1140/epjc/s10052-011-1773-4} {\bibfield  {journal}
  {\bibinfo  {journal} {Eur. Phys. J. C}\ }\textbf {\bibinfo {volume} {71}},\
  \bibinfo {pages} {1773} (\bibinfo {year} {2011})},\ \Eprint
  {https://arxiv.org/abs/1106.0645} {arXiv:1106.0645 [astro-ph.CO]}
  \BibitemShut {NoStop}%
\bibitem [{\citenamefont {Okumura}\ and\ \citenamefont
  {Yonezawa}(2003)}]{OKUMURA2003207}%
  \BibitemOpen
  \bibfield  {author} {\bibinfo {author} {\bibfnamefont {H.}~\bibnamefont
  {Okumura}}\ and\ \bibinfo {author} {\bibfnamefont {F.}~\bibnamefont
  {Yonezawa}},\ }\bibfield  {title} {\bibinfo {title} {New expression of the
  bulk viscosity},\ }\href
  {https://doi.org/https://doi.org/10.1016/S0378-4371(02)01799-5} {\bibfield
  {journal} {\bibinfo  {journal} {Physica A: Statistical Mechanics and its
  Applications}\ }\textbf {\bibinfo {volume} {321}},\ \bibinfo {pages} {207}
  (\bibinfo {year} {2003})},\ \bibinfo {note} {statphys-Taiwan-2002: Lattice
  Models and Complex Systems}\BibitemShut {NoStop}%
\bibitem [{\citenamefont {Hu}\ and\ \citenamefont {Meng}(2006)}]{Hu:2005fu}%
  \BibitemOpen
  \bibfield  {author} {\bibinfo {author} {\bibfnamefont {M.-G.}\ \bibnamefont
  {Hu}}\ and\ \bibinfo {author} {\bibfnamefont {X.-H.}\ \bibnamefont {Meng}},\
  }\bibfield  {title} {\bibinfo {title} {{Bulk viscous cosmology: statefinder
  and entropy}},\ }\href {https://doi.org/10.1016/j.physletb.2006.02.059}
  {\bibfield  {journal} {\bibinfo  {journal} {Phys. Lett. B}\ }\textbf
  {\bibinfo {volume} {635}},\ \bibinfo {pages} {186} (\bibinfo {year}
  {2006})},\ \Eprint {https://arxiv.org/abs/astro-ph/0511615}
  {arXiv:astro-ph/0511615} \BibitemShut {NoStop}%
\bibitem [{\citenamefont {Feng}\ and\ \citenamefont {Li}(2009)}]{Feng:2009jr}%
  \BibitemOpen
  \bibfield  {author} {\bibinfo {author} {\bibfnamefont {C.-J.}\ \bibnamefont
  {Feng}}\ and\ \bibinfo {author} {\bibfnamefont {X.-Z.}\ \bibnamefont {Li}},\
  }\bibfield  {title} {\bibinfo {title} {{Viscous Ricci Dark Energy}},\ }\href
  {https://doi.org/10.1016/j.physletb.2009.09.013} {\bibfield  {journal}
  {\bibinfo  {journal} {Phys. Lett. B}\ }\textbf {\bibinfo {volume} {680}},\
  \bibinfo {pages} {355} (\bibinfo {year} {2009})},\ \Eprint
  {https://arxiv.org/abs/0905.0527} {arXiv:0905.0527 [astro-ph.CO]}
  \BibitemShut {NoStop}%
\bibitem [{\citenamefont {Setare}\ and\ \citenamefont
  {Sheykhi}(2010)}]{Setare:2010zz}%
  \BibitemOpen
  \bibfield  {author} {\bibinfo {author} {\bibfnamefont {M.~R.}\ \bibnamefont
  {Setare}}\ and\ \bibinfo {author} {\bibfnamefont {A.}~\bibnamefont
  {Sheykhi}},\ }\bibfield  {title} {\bibinfo {title} {{Viscous dark energy and
  generalized second law of thermodynamics}},\ }\href
  {https://doi.org/10.1142/S0218271810017202} {\bibfield  {journal} {\bibinfo
  {journal} {Int. J. Mod. Phys. D}\ }\textbf {\bibinfo {volume} {19}},\
  \bibinfo {pages} {1205} (\bibinfo {year} {2010})},\ \Eprint
  {https://arxiv.org/abs/1103.1067} {arXiv:1103.1067 [physics.gen-ph]}
  \BibitemShut {NoStop}%
\bibitem [{\citenamefont {Hip\'olito-Ricaldi}\ \emph
  {et~al.}(2010)\citenamefont {Hip\'olito-Ricaldi}, \citenamefont {Velten},\
  and\ \citenamefont {Zimdahl}}]{PhysRevD.82.063507}%
  \BibitemOpen
  \bibfield  {author} {\bibinfo {author} {\bibfnamefont {W.~S.}\ \bibnamefont
  {Hip\'olito-Ricaldi}}, \bibinfo {author} {\bibfnamefont {H.~E.~S.}\
  \bibnamefont {Velten}},\ and\ \bibinfo {author} {\bibfnamefont
  {W.}~\bibnamefont {Zimdahl}},\ }\bibfield  {title} {\bibinfo {title} {Viscous
  dark fluid universe},\ }\href {https://doi.org/10.1103/PhysRevD.82.063507}
  {\bibfield  {journal} {\bibinfo  {journal} {Phys. Rev. D}\ }\textbf {\bibinfo
  {volume} {82}},\ \bibinfo {pages} {063507} (\bibinfo {year}
  {2010})}\BibitemShut {NoStop}%
\bibitem [{\citenamefont {Montiel}\ and\ \citenamefont
  {Breton}(2011)}]{Montiel:2011gw}%
  \BibitemOpen
  \bibfield  {author} {\bibinfo {author} {\bibfnamefont {A.}~\bibnamefont
  {Montiel}}\ and\ \bibinfo {author} {\bibfnamefont {N.}~\bibnamefont
  {Breton}},\ }\bibfield  {title} {\bibinfo {title} {{Probing bulk viscous
  matter-dominated models with Gamma-ray bursts}},\ }\href
  {https://doi.org/10.1088/1475-7516/2011/08/023} {\bibfield  {journal}
  {\bibinfo  {journal} {JCAP}\ }\textbf {\bibinfo {volume} {08}},\ \bibinfo
  {pages} {023}},\ \Eprint {https://arxiv.org/abs/1107.0271} {arXiv:1107.0271
  [astro-ph.CO]} \BibitemShut {NoStop}%
\bibitem [{\citenamefont {Wang}\ \emph {et~al.}(2017)\citenamefont {Wang},
  \citenamefont {Yan},\ and\ \citenamefont {Meng}}]{Wang:2017klo}%
  \BibitemOpen
  \bibfield  {author} {\bibinfo {author} {\bibfnamefont {D.}~\bibnamefont
  {Wang}}, \bibinfo {author} {\bibfnamefont {Y.-J.}\ \bibnamefont {Yan}},\ and\
  \bibinfo {author} {\bibfnamefont {X.-H.}\ \bibnamefont {Meng}},\ }\bibfield
  {title} {\bibinfo {title} {{Constraining viscous dark energy models with the
  latest cosmological data}},\ }\href
  {https://doi.org/10.1140/epjc/s10052-017-5212-z} {\bibfield  {journal}
  {\bibinfo  {journal} {Eur. Phys. J. C}\ }\textbf {\bibinfo {volume} {77}},\
  \bibinfo {pages} {660} (\bibinfo {year} {2017})},\ \Eprint
  {https://arxiv.org/abs/2103.14788} {arXiv:2103.14788 [astro-ph.CO]}
  \BibitemShut {NoStop}%
\bibitem [{\citenamefont {Brevik}\ \emph {et~al.}(2017)\citenamefont {Brevik},
  \citenamefont {Gr\o{}n}, \citenamefont {de~Haro}, \citenamefont {Odintsov},\
  and\ \citenamefont {Saridakis}}]{Brevik:2017msy}%
  \BibitemOpen
  \bibfield  {author} {\bibinfo {author} {\bibfnamefont {I.}~\bibnamefont
  {Brevik}}, \bibinfo {author} {\bibfnamefont {O.}~\bibnamefont {Gr\o{}n}},
  \bibinfo {author} {\bibfnamefont {J.}~\bibnamefont {de~Haro}}, \bibinfo
  {author} {\bibfnamefont {S.~D.}\ \bibnamefont {Odintsov}},\ and\ \bibinfo
  {author} {\bibfnamefont {E.~N.}\ \bibnamefont {Saridakis}},\ }\bibfield
  {title} {\bibinfo {title} {{Viscous Cosmology for Early- and Late-Time
  Universe}},\ }\href {https://doi.org/10.1142/S0218271817300245} {\bibfield
  {journal} {\bibinfo  {journal} {Int. J. Mod. Phys. D}\ }\textbf {\bibinfo
  {volume} {26}},\ \bibinfo {pages} {1730024} (\bibinfo {year} {2017})},\
  \Eprint {https://arxiv.org/abs/1706.02543} {arXiv:1706.02543 [gr-qc]}
  \BibitemShut {NoStop}%
\bibitem [{\citenamefont {da~Silva}\ and\ \citenamefont
  {Silva}(2021)}]{daSilva:2020mvk}%
  \BibitemOpen
  \bibfield  {author} {\bibinfo {author} {\bibfnamefont {W.~J.~C.}\
  \bibnamefont {da~Silva}}\ and\ \bibinfo {author} {\bibfnamefont
  {R.}~\bibnamefont {Silva}},\ }\bibfield  {title} {\bibinfo {title} {{Growth
  of matter perturbations in the extended viscous dark energy models}},\ }\href
  {https://doi.org/10.1140/epjc/s10052-021-09177-7} {\bibfield  {journal}
  {\bibinfo  {journal} {Eur. Phys. J. C}\ }\textbf {\bibinfo {volume} {81}},\
  \bibinfo {pages} {403} (\bibinfo {year} {2021})},\ \Eprint
  {https://arxiv.org/abs/2011.09516} {arXiv:2011.09516 [astro-ph.CO]}
  \BibitemShut {NoStop}%
\bibitem [{\citenamefont {Velten}\ and\ \citenamefont
  {Schwarz}(2012)}]{PhysRevD.86.083501}%
  \BibitemOpen
  \bibfield  {author} {\bibinfo {author} {\bibfnamefont {H.}~\bibnamefont
  {Velten}}\ and\ \bibinfo {author} {\bibfnamefont {D.~J.}\ \bibnamefont
  {Schwarz}},\ }\bibfield  {title} {\bibinfo {title} {Dissipation of dark
  matter},\ }\href {https://doi.org/10.1103/PhysRevD.86.083501} {\bibfield
  {journal} {\bibinfo  {journal} {Phys. Rev. D}\ }\textbf {\bibinfo {volume}
  {86}},\ \bibinfo {pages} {083501} (\bibinfo {year} {2012})}\BibitemShut
  {NoStop}%
\bibitem [{\citenamefont {Brevik}(2012)}]{Brevik:2012bc}%
  \BibitemOpen
  \bibfield  {author} {\bibinfo {author} {\bibfnamefont {I.}~\bibnamefont
  {Brevik}},\ }\bibfield  {title} {\bibinfo {title} {{Viscosity in modified
  gravity}},\ }\href {https://doi.org/10.3390/e14112302} {\bibfield  {journal}
  {\bibinfo  {journal} {Entropy}\ }\textbf {\bibinfo {volume} {14}},\ \bibinfo
  {pages} {2302} (\bibinfo {year} {2012})},\ \Eprint
  {https://arxiv.org/abs/1211.2562} {arXiv:1211.2562 [gr-qc]} \BibitemShut
  {NoStop}%
\bibitem [{\citenamefont {Singh}\ and\ \citenamefont
  {Kumar}(2014)}]{Singh:2014bha}%
  \BibitemOpen
  \bibfield  {author} {\bibinfo {author} {\bibfnamefont {C.~P.}\ \bibnamefont
  {Singh}}\ and\ \bibinfo {author} {\bibfnamefont {P.}~\bibnamefont {Kumar}},\
  }\bibfield  {title} {\bibinfo {title} {{Friedmann model with viscous
  cosmology in modified $f(R,T)$ gravity theory}},\ }\href
  {https://doi.org/10.1140/epjc/s10052-014-3070-5} {\bibfield  {journal}
  {\bibinfo  {journal} {Eur. Phys. J. C}\ }\textbf {\bibinfo {volume} {74}},\
  \bibinfo {pages} {3070} (\bibinfo {year} {2014})},\ \Eprint
  {https://arxiv.org/abs/1406.4258} {arXiv:1406.4258 [gr-qc]} \BibitemShut
  {NoStop}%
\bibitem [{\citenamefont {Velten}\ \emph {et~al.}(2013)\citenamefont {Velten},
  \citenamefont {Wang},\ and\ \citenamefont {Meng}}]{Velten:2013qna}%
  \BibitemOpen
  \bibfield  {author} {\bibinfo {author} {\bibfnamefont {H.}~\bibnamefont
  {Velten}}, \bibinfo {author} {\bibfnamefont {J.}~\bibnamefont {Wang}},\ and\
  \bibinfo {author} {\bibfnamefont {X.}~\bibnamefont {Meng}},\ }\bibfield
  {title} {\bibinfo {title} {{Phantom dark energy as an effect of bulk
  viscosity}},\ }\href {https://doi.org/10.1103/PhysRevD.88.123504} {\bibfield
  {journal} {\bibinfo  {journal} {Phys. Rev. D}\ }\textbf {\bibinfo {volume}
  {88}},\ \bibinfo {pages} {123504} (\bibinfo {year} {2013})},\ \Eprint
  {https://arxiv.org/abs/1307.4262} {arXiv:1307.4262 [astro-ph.CO]}
  \BibitemShut {NoStop}%
\bibitem [{\citenamefont {Hu}\ and\ \citenamefont {Hu}(2020)}]{Hu:2020xus}%
  \BibitemOpen
  \bibfield  {author} {\bibinfo {author} {\bibfnamefont {J.}~\bibnamefont
  {Hu}}\ and\ \bibinfo {author} {\bibfnamefont {H.}~\bibnamefont {Hu}},\
  }\bibfield  {title} {\bibinfo {title} {{Viscous universe with cosmological
  constant}},\ }\href {https://doi.org/10.1140/epjp/s13360-020-00623-1}
  {\bibfield  {journal} {\bibinfo  {journal} {Eur. Phys. J. Plus}\ }\textbf
  {\bibinfo {volume} {135}},\ \bibinfo {pages} {718} (\bibinfo {year}
  {2020})}\BibitemShut {NoStop}%
\bibitem [{\citenamefont {Arora}\ \emph {et~al.}(2020)\citenamefont {Arora},
  \citenamefont {Meng}, \citenamefont {Pacif},\ and\ \citenamefont
  {Sahoo}}]{Arora:2020lsr}%
  \BibitemOpen
  \bibfield  {author} {\bibinfo {author} {\bibfnamefont {S.}~\bibnamefont
  {Arora}}, \bibinfo {author} {\bibfnamefont {X.-h.}\ \bibnamefont {Meng}},
  \bibinfo {author} {\bibfnamefont {S.~K.~J.}\ \bibnamefont {Pacif}},\ and\
  \bibinfo {author} {\bibfnamefont {P.~K.}\ \bibnamefont {Sahoo}},\ }\bibfield
  {title} {\bibinfo {title} {{Effective equation of state in modified gravity
  and observational constraints}},\ }\href
  {https://doi.org/10.1088/1361-6382/aba587} {\bibfield  {journal} {\bibinfo
  {journal} {Class. Quant. Grav.}\ }\textbf {\bibinfo {volume} {37}},\ \bibinfo
  {pages} {205022} (\bibinfo {year} {2020})},\ \Eprint
  {https://arxiv.org/abs/2007.07717} {arXiv:2007.07717 [gr-qc]} \BibitemShut
  {NoStop}%
\bibitem [{\citenamefont {Sharif}\ and\ \citenamefont
  {Rani}(2013)}]{Sharif:2013tny}%
  \BibitemOpen
  \bibfield  {author} {\bibinfo {author} {\bibfnamefont {M.}~\bibnamefont
  {Sharif}}\ and\ \bibinfo {author} {\bibfnamefont {S.}~\bibnamefont {Rani}},\
  }\bibfield  {title} {\bibinfo {title} {{Viscous Dark Energy in $f(T)$
  Gravity}},\ }\href {https://doi.org/10.1142/S0217732313501186} {\bibfield
  {journal} {\bibinfo  {journal} {Mod. Phys. Lett. A}\ }\textbf {\bibinfo
  {volume} {28}},\ \bibinfo {pages} {1350118} (\bibinfo {year} {2013})},\
  \Eprint {https://arxiv.org/abs/1405.5232} {arXiv:1405.5232 [gr-qc]}
  \BibitemShut {NoStop}%
\bibitem [{\citenamefont {Jawad}\ \emph {et~al.}(2016)\citenamefont {Jawad},
  \citenamefont {Chattopadhyay},\ and\ \citenamefont {Rani}}]{Jawad:2016omn}%
  \BibitemOpen
  \bibfield  {author} {\bibinfo {author} {\bibfnamefont {A.}~\bibnamefont
  {Jawad}}, \bibinfo {author} {\bibfnamefont {S.}~\bibnamefont
  {Chattopadhyay}},\ and\ \bibinfo {author} {\bibfnamefont {S.}~\bibnamefont
  {Rani}},\ }\bibfield  {title} {\bibinfo {title} {{Viscous pilgrim $f(T)$
  gravity models}},\ }\href {https://doi.org/10.1007/s10509-016-2814-0}
  {\bibfield  {journal} {\bibinfo  {journal} {Astrophys. Space Sci.}\ }\textbf
  {\bibinfo {volume} {361}},\ \bibinfo {pages} {231} (\bibinfo {year}
  {2016})}\BibitemShut {NoStop}%
\bibitem [{\citenamefont {Davood~Sadatian}(2019)}]{DavoodSadatian:2019pvq}%
  \BibitemOpen
  \bibfield  {author} {\bibinfo {author} {\bibfnamefont {S.}~\bibnamefont
  {Davood~Sadatian}},\ }\bibfield  {title} {\bibinfo {title} {{Effects of
  viscous content on the modified cosmological F(T) model}},\ }\href
  {https://doi.org/10.1209/0295-5075/126/30004} {\bibfield  {journal} {\bibinfo
   {journal} {EPL}\ }\textbf {\bibinfo {volume} {126}},\ \bibinfo {pages}
  {30004} (\bibinfo {year} {2019})}\BibitemShut {NoStop}%
\bibitem [{\citenamefont {Gadbail}\ \emph {et~al.}(2021)\citenamefont
  {Gadbail}, \citenamefont {Arora},\ and\ \citenamefont
  {Sahoo}}]{Gadbail:2021fjf}%
  \BibitemOpen
  \bibfield  {author} {\bibinfo {author} {\bibfnamefont {G.~N.}\ \bibnamefont
  {Gadbail}}, \bibinfo {author} {\bibfnamefont {S.}~\bibnamefont {Arora}},\
  and\ \bibinfo {author} {\bibfnamefont {P.~K.}\ \bibnamefont {Sahoo}},\
  }\bibfield  {title} {\bibinfo {title} {{Viscous cosmology in the Weyl-type
  f(Q,~T) gravity}},\ }\href {https://doi.org/10.1140/epjc/s10052-021-09889-w}
  {\bibfield  {journal} {\bibinfo  {journal} {Eur. Phys. J. C}\ }\textbf
  {\bibinfo {volume} {81}},\ \bibinfo {pages} {1088} (\bibinfo {year}
  {2021})},\ \Eprint {https://arxiv.org/abs/2110.02726} {arXiv:2110.02726
  [gr-qc]} \BibitemShut {NoStop}%
\bibitem [{\citenamefont {Arora}\ \emph {et~al.}(2022)\citenamefont {Arora},
  \citenamefont {Pacif}, \citenamefont {Parida},\ and\ \citenamefont
  {Sahoo}}]{Arora:2021tuh}%
  \BibitemOpen
  \bibfield  {author} {\bibinfo {author} {\bibfnamefont {S.}~\bibnamefont
  {Arora}}, \bibinfo {author} {\bibfnamefont {S.~K.~J.}\ \bibnamefont {Pacif}},
  \bibinfo {author} {\bibfnamefont {A.}~\bibnamefont {Parida}},\ and\ \bibinfo
  {author} {\bibfnamefont {P.~K.}\ \bibnamefont {Sahoo}},\ }\bibfield  {title}
  {\bibinfo {title} {{Bulk viscous matter and the cosmic acceleration of the
  universe in f(Q,T) gravity}},\ }\href
  {https://doi.org/10.1016/j.jheap.2021.10.001} {\bibfield  {journal} {\bibinfo
   {journal} {JHEAp}\ }\textbf {\bibinfo {volume} {33}},\ \bibinfo {pages}
  {129} (\bibinfo {year} {2022})},\ \Eprint {https://arxiv.org/abs/2106.00491}
  {arXiv:2106.00491 [gr-qc]} \BibitemShut {NoStop}%
\bibitem [{\citenamefont {Saha}\ and\ \citenamefont
  {Chattopadhyay}(2022)}]{Saha:2022ian}%
  \BibitemOpen
  \bibfield  {author} {\bibinfo {author} {\bibfnamefont {S.}~\bibnamefont
  {Saha}}\ and\ \bibinfo {author} {\bibfnamefont {S.}~\bibnamefont
  {Chattopadhyay}},\ }\bibfield  {title} {\bibinfo {title} {{Viscous
  generalised Chaplygin gas under the purview of f(T) gravity and the model
  assessment through probabilistic information theory}},\ }\href
  {https://doi.org/10.1088/1402-4896/ac5af4} {\bibfield  {journal} {\bibinfo
  {journal} {Phys. Scripta}\ }\textbf {\bibinfo {volume} {97}},\ \bibinfo
  {pages} {045006} (\bibinfo {year} {2022})}\BibitemShut {NoStop}%
\bibitem [{\citenamefont {Freedman}(2017)}]{Freedman:2017yms}%
  \BibitemOpen
  \bibfield  {author} {\bibinfo {author} {\bibfnamefont {W.~L.}\ \bibnamefont
  {Freedman}},\ }\bibfield  {title} {\bibinfo {title} {{Cosmology at a
  Crossroads}},\ }\href {https://doi.org/10.1038/s41550-017-0121} {\bibfield
  {journal} {\bibinfo  {journal} {Nature Astron.}\ }\textbf {\bibinfo {volume}
  {1}},\ \bibinfo {pages} {0121} (\bibinfo {year} {2017})},\ \Eprint
  {https://arxiv.org/abs/1706.02739} {arXiv:1706.02739 [astro-ph.CO]}
  \BibitemShut {NoStop}%
\bibitem [{\citenamefont {Evslin}\ \emph {et~al.}(2018)\citenamefont {Evslin},
  \citenamefont {Sen},\ and\ \citenamefont {Ruchika}}]{PhysRevD.97.103511}%
  \BibitemOpen
  \bibfield  {author} {\bibinfo {author} {\bibfnamefont {J.}~\bibnamefont
  {Evslin}}, \bibinfo {author} {\bibfnamefont {A.~A.}\ \bibnamefont {Sen}},\
  and\ \bibinfo {author} {\bibnamefont {Ruchika}},\ }\bibfield  {title}
  {\bibinfo {title} {Price of shifting the hubble constant},\ }\href
  {https://doi.org/10.1103/PhysRevD.97.103511} {\bibfield  {journal} {\bibinfo
  {journal} {Phys. Rev. D}\ }\textbf {\bibinfo {volume} {97}},\ \bibinfo
  {pages} {103511} (\bibinfo {year} {2018})}\BibitemShut {NoStop}%
\bibitem [{\citenamefont {Aylor}\ \emph {et~al.}(2019)\citenamefont {Aylor},
  \citenamefont {Joy}, \citenamefont {Knox}, \citenamefont {Millea},
  \citenamefont {Raghunathan},\ and\ \citenamefont {Wu}}]{Aylor:2018drw}%
  \BibitemOpen
  \bibfield  {author} {\bibinfo {author} {\bibfnamefont {K.}~\bibnamefont
  {Aylor}}, \bibinfo {author} {\bibfnamefont {M.}~\bibnamefont {Joy}}, \bibinfo
  {author} {\bibfnamefont {L.}~\bibnamefont {Knox}}, \bibinfo {author}
  {\bibfnamefont {M.}~\bibnamefont {Millea}}, \bibinfo {author} {\bibfnamefont
  {S.}~\bibnamefont {Raghunathan}},\ and\ \bibinfo {author} {\bibfnamefont
  {W.~L.~K.}\ \bibnamefont {Wu}},\ }\bibfield  {title} {\bibinfo {title}
  {{Sounds Discordant: Classical Distance Ladder \& $\Lambda$CDM -based
  Determinations of the Cosmological Sound Horizon}},\ }\href
  {https://doi.org/10.3847/1538-4357/ab0898} {\bibfield  {journal} {\bibinfo
  {journal} {Astrophys. J.}\ }\textbf {\bibinfo {volume} {874}},\ \bibinfo
  {pages} {4} (\bibinfo {year} {2019})},\ \Eprint
  {https://arxiv.org/abs/1811.00537} {arXiv:1811.00537 [astro-ph.CO]}
  \BibitemShut {NoStop}%
\bibitem [{\citenamefont {Verde}\ \emph {et~al.}(2019)\citenamefont {Verde},
  \citenamefont {Treu},\ and\ \citenamefont {Riess}}]{Verde:2019ivm}%
  \BibitemOpen
  \bibfield  {author} {\bibinfo {author} {\bibfnamefont {L.}~\bibnamefont
  {Verde}}, \bibinfo {author} {\bibfnamefont {T.}~\bibnamefont {Treu}},\ and\
  \bibinfo {author} {\bibfnamefont {A.~G.}\ \bibnamefont {Riess}},\ }\bibfield
  {title} {\bibinfo {title} {{Tensions between the Early and the Late
  Universe}},\ }\href {https://doi.org/10.1038/s41550-019-0902-0} {\bibfield
  {journal} {\bibinfo  {journal} {Nature Astron.}\ }\textbf {\bibinfo {volume}
  {3}},\ \bibinfo {pages} {891} (\bibinfo {year} {2019})},\ \Eprint
  {https://arxiv.org/abs/1907.10625} {arXiv:1907.10625 [astro-ph.CO]}
  \BibitemShut {NoStop}%
\bibitem [{\citenamefont {Ren}\ and\ \citenamefont
  {Meng}(2006{\natexlab{b}})}]{Ren:2005nw}%
  \BibitemOpen
  \bibfield  {author} {\bibinfo {author} {\bibfnamefont {J.}~\bibnamefont
  {Ren}}\ and\ \bibinfo {author} {\bibfnamefont {X.-H.}\ \bibnamefont {Meng}},\
  }\bibfield  {title} {\bibinfo {title} {{Cosmological model with viscosity
  media (dark fluid) described by an effective equation of state}},\ }\href
  {https://doi.org/10.1016/j.physletb.2005.11.055} {\bibfield  {journal}
  {\bibinfo  {journal} {Phys. Lett. B}\ }\textbf {\bibinfo {volume} {633}},\
  \bibinfo {pages} {1} (\bibinfo {year} {2006}{\natexlab{b}})},\ \Eprint
  {https://arxiv.org/abs/astro-ph/0511163} {arXiv:astro-ph/0511163}
  \BibitemShut {NoStop}%
\bibitem [{\citenamefont {Sasidharan}\ \emph {et~al.}(2018)\citenamefont
  {Sasidharan}, \citenamefont {Mohan}, \citenamefont {John},\ and\
  \citenamefont {Mathew}}]{Sasidharan:2018bay}%
  \BibitemOpen
  \bibfield  {author} {\bibinfo {author} {\bibfnamefont {A.}~\bibnamefont
  {Sasidharan}}, \bibinfo {author} {\bibfnamefont {N.~D.~J.}\ \bibnamefont
  {Mohan}}, \bibinfo {author} {\bibfnamefont {M.~V.}\ \bibnamefont {John}},\
  and\ \bibinfo {author} {\bibfnamefont {T.~K.}\ \bibnamefont {Mathew}},\
  }\bibfield  {title} {\bibinfo {title} {{Bayesian analysis of bulk viscous
  matter dominated universe}},\ }\href
  {https://doi.org/10.1140/epjc/s10052-018-6105-5} {\bibfield  {journal}
  {\bibinfo  {journal} {Eur. Phys. J. C}\ }\textbf {\bibinfo {volume} {78}},\
  \bibinfo {pages} {628} (\bibinfo {year} {2018})},\ \Eprint
  {https://arxiv.org/abs/1803.08235} {arXiv:1803.08235 [gr-qc]} \BibitemShut
  {NoStop}%
\bibitem [{\citenamefont {Capozziello}\ \emph {et~al.}(2006)\citenamefont
  {Capozziello}, \citenamefont {Cardone}, \citenamefont {Elizalde},
  \citenamefont {Nojiri},\ and\ \citenamefont {Odintsov}}]{PhysRevD.73.043512}%
  \BibitemOpen
  \bibfield  {author} {\bibinfo {author} {\bibfnamefont {S.}~\bibnamefont
  {Capozziello}}, \bibinfo {author} {\bibfnamefont {V.~F.}\ \bibnamefont
  {Cardone}}, \bibinfo {author} {\bibfnamefont {E.}~\bibnamefont {Elizalde}},
  \bibinfo {author} {\bibfnamefont {S.}~\bibnamefont {Nojiri}},\ and\ \bibinfo
  {author} {\bibfnamefont {S.~D.}\ \bibnamefont {Odintsov}},\ }\bibfield
  {title} {\bibinfo {title} {Observational constraints on dark energy with
  generalized equations of state},\ }\href
  {https://doi.org/10.1103/PhysRevD.73.043512} {\bibfield  {journal} {\bibinfo
  {journal} {Phys. Rev. D}\ }\textbf {\bibinfo {volume} {73}},\ \bibinfo
  {pages} {043512} (\bibinfo {year} {2006})}\BibitemShut {NoStop}%
\bibitem [{\citenamefont {Odintsov}\ \emph {et~al.}(2020)\citenamefont
  {Odintsov}, \citenamefont {G\'omez},\ and\ \citenamefont
  {Sharov}}]{PhysRevD.101.044010}%
  \BibitemOpen
  \bibfield  {author} {\bibinfo {author} {\bibfnamefont {S.~D.}\ \bibnamefont
  {Odintsov}}, \bibinfo {author} {\bibfnamefont {D.~S.-C.}\ \bibnamefont
  {G\'omez}},\ and\ \bibinfo {author} {\bibfnamefont {G.~S.}\ \bibnamefont
  {Sharov}},\ }\bibfield  {title} {\bibinfo {title} {Testing the equation of
  state for viscous dark energy},\ }\href
  {https://doi.org/10.1103/PhysRevD.101.044010} {\bibfield  {journal} {\bibinfo
   {journal} {Phys. Rev. D}\ }\textbf {\bibinfo {volume} {101}},\ \bibinfo
  {pages} {044010} (\bibinfo {year} {2020})}\BibitemShut {NoStop}%
\bibitem [{\citenamefont {Padmanabhan}\ and\ \citenamefont
  {Chitre}(1987)}]{Padmanabhan:1987dg}%
  \BibitemOpen
  \bibfield  {author} {\bibinfo {author} {\bibfnamefont {T.}~\bibnamefont
  {Padmanabhan}}\ and\ \bibinfo {author} {\bibfnamefont {S.~M.}\ \bibnamefont
  {Chitre}},\ }\bibfield  {title} {\bibinfo {title} {{Viscous universes}},\
  }\href {https://doi.org/10.1016/0375-9601(87)90104-6} {\bibfield  {journal}
  {\bibinfo  {journal} {Phys. Lett. A}\ }\textbf {\bibinfo {volume} {120}},\
  \bibinfo {pages} {433} (\bibinfo {year} {1987})}\BibitemShut {NoStop}%
\bibitem [{\citenamefont {Szydlowski}\ and\ \citenamefont
  {Hrycyna}(2007)}]{Szydlowski:2006ma}%
  \BibitemOpen
  \bibfield  {author} {\bibinfo {author} {\bibfnamefont {M.}~\bibnamefont
  {Szydlowski}}\ and\ \bibinfo {author} {\bibfnamefont {O.}~\bibnamefont
  {Hrycyna}},\ }\bibfield  {title} {\bibinfo {title} {{Dissipative or
  conservative cosmology with dark energy?}},\ }\href
  {https://doi.org/10.1016/j.aop.2007.06.008} {\bibfield  {journal} {\bibinfo
  {journal} {Annals Phys.}\ }\textbf {\bibinfo {volume} {322}},\ \bibinfo
  {pages} {2745} (\bibinfo {year} {2007})},\ \Eprint
  {https://arxiv.org/abs/astro-ph/0602118} {arXiv:astro-ph/0602118}
  \BibitemShut {NoStop}%
\bibitem [{\citenamefont {Hipolito-Ricaldi}\ \emph {et~al.}(2009)\citenamefont
  {Hipolito-Ricaldi}, \citenamefont {Velten},\ and\ \citenamefont
  {Zimdahl}}]{Hipolito-Ricaldi:2009xbk}%
  \BibitemOpen
  \bibfield  {author} {\bibinfo {author} {\bibfnamefont {W.~S.}\ \bibnamefont
  {Hipolito-Ricaldi}}, \bibinfo {author} {\bibfnamefont {H.~E.~S.}\
  \bibnamefont {Velten}},\ and\ \bibinfo {author} {\bibfnamefont
  {W.}~\bibnamefont {Zimdahl}},\ }\bibfield  {title} {\bibinfo {title}
  {{Non-adiabatic dark fluid cosmology}},\ }\href
  {https://doi.org/10.1088/1475-7516/2009/06/016} {\bibfield  {journal}
  {\bibinfo  {journal} {JCAP}\ }\textbf {\bibinfo {volume} {06}},\ \bibinfo
  {pages} {016}},\ \Eprint {https://arxiv.org/abs/0902.4710} {arXiv:0902.4710
  [astro-ph.CO]} \BibitemShut {NoStop}%
\bibitem [{\citenamefont {Mohan}\ \emph {et~al.}(2017)\citenamefont {Mohan},
  \citenamefont {Sasidharan},\ and\ \citenamefont {Mathew}}]{Mohan:2017poq}%
  \BibitemOpen
  \bibfield  {author} {\bibinfo {author} {\bibfnamefont {N.~D.~J.}\
  \bibnamefont {Mohan}}, \bibinfo {author} {\bibfnamefont {A.}~\bibnamefont
  {Sasidharan}},\ and\ \bibinfo {author} {\bibfnamefont {T.~K.}\ \bibnamefont
  {Mathew}},\ }\bibfield  {title} {\bibinfo {title} {{Bulk viscous matter and
  recent acceleration of the universe based on causal viscous theory}},\ }\href
  {https://doi.org/10.1140/epjc/s10052-017-5428-y} {\bibfield  {journal}
  {\bibinfo  {journal} {Eur. Phys. J. C}\ }\textbf {\bibinfo {volume} {77}},\
  \bibinfo {pages} {849} (\bibinfo {year} {2017})},\ \Eprint
  {https://arxiv.org/abs/1708.02437} {arXiv:1708.02437 [gr-qc]} \BibitemShut
  {NoStop}%
\bibitem [{\citenamefont {Feng}\ \emph {et~al.}(2015)\citenamefont {Feng},
  \citenamefont {Ge}, \citenamefont {Li}, \citenamefont {Lin},\ and\
  \citenamefont {Zhai}}]{Feng:2015awr}%
  \BibitemOpen
  \bibfield  {author} {\bibinfo {author} {\bibfnamefont {C.-j.}\ \bibnamefont
  {Feng}}, \bibinfo {author} {\bibfnamefont {F.-f.}\ \bibnamefont {Ge}},
  \bibinfo {author} {\bibfnamefont {X.-z.}\ \bibnamefont {Li}}, \bibinfo
  {author} {\bibfnamefont {R.-h.}\ \bibnamefont {Lin}},\ and\ \bibinfo {author}
  {\bibfnamefont {X.-h.}\ \bibnamefont {Zhai}},\ }\bibfield  {title} {\bibinfo
  {title} {{Towards realistic $f(T)$ models with nonminimal torsion-matter
  coupling extension}},\ }\href {https://doi.org/10.1103/PhysRevD.92.104038}
  {\bibfield  {journal} {\bibinfo  {journal} {Phys. Rev. D}\ }\textbf {\bibinfo
  {volume} {92}},\ \bibinfo {pages} {104038} (\bibinfo {year} {2015})},\
  \Eprint {https://arxiv.org/abs/1511.07935} {arXiv:1511.07935 [gr-qc]}
  \BibitemShut {NoStop}%
\bibitem [{\citenamefont {Lin}\ \emph {et~al.}(2017)\citenamefont {Lin},
  \citenamefont {Zhai},\ and\ \citenamefont {Li}}]{Lin:2016nvj}%
  \BibitemOpen
  \bibfield  {author} {\bibinfo {author} {\bibfnamefont {R.-H.}\ \bibnamefont
  {Lin}}, \bibinfo {author} {\bibfnamefont {X.-H.}\ \bibnamefont {Zhai}},\ and\
  \bibinfo {author} {\bibfnamefont {X.-Z.}\ \bibnamefont {Li}},\ }\bibfield
  {title} {\bibinfo {title} {{Solar system tests for realistic $f(T)$ models
  with non-minimal torsion\textendash{}matter coupling}},\ }\href
  {https://doi.org/10.1140/epjc/s10052-017-5074-4} {\bibfield  {journal}
  {\bibinfo  {journal} {Eur. Phys. J. C}\ }\textbf {\bibinfo {volume} {77}},\
  \bibinfo {pages} {504} (\bibinfo {year} {2017})},\ \Eprint
  {https://arxiv.org/abs/1610.04956} {arXiv:1610.04956 [gr-qc]} \BibitemShut
  {NoStop}%
\bibitem [{\citenamefont {Lin}\ \emph {et~al.}(2018)\citenamefont {Lin},
  \citenamefont {Wen}, \citenamefont {Zhai},\ and\ \citenamefont
  {Li}}]{Lin:2018xzd}%
  \BibitemOpen
  \bibfield  {author} {\bibinfo {author} {\bibfnamefont {R.-H.}\ \bibnamefont
  {Lin}}, \bibinfo {author} {\bibfnamefont {Q.}~\bibnamefont {Wen}}, \bibinfo
  {author} {\bibfnamefont {X.-H.}\ \bibnamefont {Zhai}},\ and\ \bibinfo
  {author} {\bibfnamefont {X.-Z.}\ \bibnamefont {Li}},\ }\bibfield  {title}
  {\bibinfo {title} {{Diagnostics for generalized power-law
  torsion\textendash{}matter coupling $f(T)$ model}},\ }\href
  {https://doi.org/10.1142/S0218271819500317} {\bibfield  {journal} {\bibinfo
  {journal} {Int. J. Mod. Phys. D}\ }\textbf {\bibinfo {volume} {28}},\
  \bibinfo {pages} {1950031} (\bibinfo {year} {2018})},\ \Eprint
  {https://arxiv.org/abs/1803.04134} {arXiv:1803.04134 [gr-qc]} \BibitemShut
  {NoStop}%
\bibitem [{\citenamefont {Scolnic}\ \emph {et~al.}(2018)\citenamefont {Scolnic}
  \emph {et~al.}}]{Pan-STARRS1:2017jku}%
  \BibitemOpen
  \bibfield  {author} {\bibinfo {author} {\bibfnamefont {D.~M.}\ \bibnamefont
  {Scolnic}} \emph {et~al.} (\bibinfo {collaboration} {Pan-STARRS1}),\
  }\bibfield  {title} {\bibinfo {title} {{The Complete Light-curve Sample of
  Spectroscopically Confirmed SNe Ia from Pan-STARRS1 and Cosmological
  Constraints from the Combined Pantheon Sample}},\ }\href
  {https://doi.org/10.3847/1538-4357/aab9bb} {\bibfield  {journal} {\bibinfo
  {journal} {Astrophys. J.}\ }\textbf {\bibinfo {volume} {859}},\ \bibinfo
  {pages} {101} (\bibinfo {year} {2018})},\ \Eprint
  {https://arxiv.org/abs/1710.00845} {arXiv:1710.00845 [astro-ph.CO]}
  \BibitemShut {NoStop}%
\bibitem [{\citenamefont {Sharov}\ and\ \citenamefont
  {Vasiliev}(2018)}]{Sharov:2018yvz}%
  \BibitemOpen
  \bibfield  {author} {\bibinfo {author} {\bibfnamefont {G.~S.}\ \bibnamefont
  {Sharov}}\ and\ \bibinfo {author} {\bibfnamefont {V.~O.}\ \bibnamefont
  {Vasiliev}},\ }\bibfield  {title} {\bibinfo {title} {{How predictions of
  cosmological models depend on Hubble parameter data sets}},\ }\href
  {https://doi.org/10.26456/mmg/2018-611} {\bibfield  {journal} {\bibinfo
  {journal} {Math. Model. Geom.}\ }\textbf {\bibinfo {volume} {6}},\ \bibinfo
  {pages} {1} (\bibinfo {year} {2018})},\ \Eprint
  {https://arxiv.org/abs/1807.07323} {arXiv:1807.07323 [gr-qc]} \BibitemShut
  {NoStop}%
\bibitem [{\citenamefont {Betoule}\ \emph {et~al.}(2014)\citenamefont {Betoule}
  \emph {et~al.}}]{SDSS:2014iwm}%
  \BibitemOpen
  \bibfield  {author} {\bibinfo {author} {\bibfnamefont {M.}~\bibnamefont
  {Betoule}} \emph {et~al.} (\bibinfo {collaboration} {SDSS}),\ }\bibfield
  {title} {\bibinfo {title} {{Improved cosmological constraints from a joint
  analysis of the SDSS-II and SNLS supernova samples}},\ }\href
  {https://doi.org/10.1051/0004-6361/201423413} {\bibfield  {journal} {\bibinfo
   {journal} {Astron. Astrophys.}\ }\textbf {\bibinfo {volume} {568}},\
  \bibinfo {pages} {A22} (\bibinfo {year} {2014})},\ \Eprint
  {https://arxiv.org/abs/1401.4064} {arXiv:1401.4064 [astro-ph.CO]}
  \BibitemShut {NoStop}%
\bibitem [{\citenamefont {Aghanim}\ \emph {et~al.}(2020)\citenamefont {Aghanim}
  \emph {et~al.}}]{Planck:2018vyg}%
  \BibitemOpen
  \bibfield  {author} {\bibinfo {author} {\bibfnamefont {N.}~\bibnamefont
  {Aghanim}} \emph {et~al.} (\bibinfo {collaboration} {Planck}),\ }\bibfield
  {title} {\bibinfo {title} {{Planck 2018 results. VI. Cosmological
  parameters}},\ }\href {https://doi.org/10.1051/0004-6361/201833910}
  {\bibfield  {journal} {\bibinfo  {journal} {Astron. Astrophys.}\ }\textbf
  {\bibinfo {volume} {641}},\ \bibinfo {pages} {A6} (\bibinfo {year} {2020})},\
  \bibinfo {note} {[Erratum: Astron.Astrophys. 652, C4 (2021)]},\ \Eprint
  {https://arxiv.org/abs/1807.06209} {arXiv:1807.06209 [astro-ph.CO]}
  \BibitemShut {NoStop}%
\bibitem [{\citenamefont {Beutler}\ \emph {et~al.}(2011)\citenamefont
  {Beutler}, \citenamefont {Blake}, \citenamefont {Colless}, \citenamefont
  {Jones}, \citenamefont {Staveley-Smith}, \citenamefont {Campbell},
  \citenamefont {Parker}, \citenamefont {Saunders},\ and\ \citenamefont
  {Watson}}]{Beutler:2011hx}%
  \BibitemOpen
  \bibfield  {author} {\bibinfo {author} {\bibfnamefont {F.}~\bibnamefont
  {Beutler}}, \bibinfo {author} {\bibfnamefont {C.}~\bibnamefont {Blake}},
  \bibinfo {author} {\bibfnamefont {M.}~\bibnamefont {Colless}}, \bibinfo
  {author} {\bibfnamefont {D.~H.}\ \bibnamefont {Jones}}, \bibinfo {author}
  {\bibfnamefont {L.}~\bibnamefont {Staveley-Smith}}, \bibinfo {author}
  {\bibfnamefont {L.}~\bibnamefont {Campbell}}, \bibinfo {author}
  {\bibfnamefont {Q.}~\bibnamefont {Parker}}, \bibinfo {author} {\bibfnamefont
  {W.}~\bibnamefont {Saunders}},\ and\ \bibinfo {author} {\bibfnamefont
  {F.}~\bibnamefont {Watson}},\ }\bibfield  {title} {\bibinfo {title} {{The 6dF
  Galaxy Survey: Baryon Acoustic Oscillations and the Local Hubble Constant}},\
  }\href {https://doi.org/10.1111/j.1365-2966.2011.19250.x} {\bibfield
  {journal} {\bibinfo  {journal} {Mon. Not. Roy. Astron. Soc.}\ }\textbf
  {\bibinfo {volume} {416}},\ \bibinfo {pages} {3017} (\bibinfo {year}
  {2011})},\ \Eprint {https://arxiv.org/abs/1106.3366} {arXiv:1106.3366
  [astro-ph.CO]} \BibitemShut {NoStop}%
\bibitem [{\citenamefont {Padmanabhan}\ \emph {et~al.}(2012)\citenamefont
  {Padmanabhan}, \citenamefont {Xu}, \citenamefont {Eisenstein}, \citenamefont
  {Scalzo}, \citenamefont {Cuesta}, \citenamefont {Mehta},\ and\ \citenamefont
  {Kazin}}]{Padmanabhan:2012hf}%
  \BibitemOpen
  \bibfield  {author} {\bibinfo {author} {\bibfnamefont {N.}~\bibnamefont
  {Padmanabhan}}, \bibinfo {author} {\bibfnamefont {X.}~\bibnamefont {Xu}},
  \bibinfo {author} {\bibfnamefont {D.~J.}\ \bibnamefont {Eisenstein}},
  \bibinfo {author} {\bibfnamefont {R.}~\bibnamefont {Scalzo}}, \bibinfo
  {author} {\bibfnamefont {A.~J.}\ \bibnamefont {Cuesta}}, \bibinfo {author}
  {\bibfnamefont {K.~T.}\ \bibnamefont {Mehta}},\ and\ \bibinfo {author}
  {\bibfnamefont {E.}~\bibnamefont {Kazin}},\ }\bibfield  {title} {\bibinfo
  {title} {{A 2 per cent distance to $z$=0.35 by reconstructing baryon acoustic
  oscillations - I. Methods and application to the Sloan Digital Sky Survey}},\
  }\href {https://doi.org/10.1111/j.1365-2966.2012.21888.x} {\bibfield
  {journal} {\bibinfo  {journal} {Mon. Not. Roy. Astron. Soc.}\ }\textbf
  {\bibinfo {volume} {427}},\ \bibinfo {pages} {2132} (\bibinfo {year}
  {2012})},\ \Eprint {https://arxiv.org/abs/1202.0090} {arXiv:1202.0090
  [astro-ph.CO]} \BibitemShut {NoStop}%
\bibitem [{\citenamefont {Anderson}\ \emph {et~al.}(2013)\citenamefont
  {Anderson} \emph {et~al.}}]{Anderson:2012sa}%
  \BibitemOpen
  \bibfield  {author} {\bibinfo {author} {\bibfnamefont {L.}~\bibnamefont
  {Anderson}} \emph {et~al.},\ }\bibfield  {title} {\bibinfo {title} {{The
  clustering of galaxies in the SDSS-III Baryon Oscillation Spectroscopic
  Survey: Baryon Acoustic Oscillations in the Data Release 9 Spectroscopic
  Galaxy Sample}},\ }\href {https://doi.org/10.1111/j.1365-2966.2012.22066.x}
  {\bibfield  {journal} {\bibinfo  {journal} {Mon. Not. Roy. Astron. Soc.}\
  }\textbf {\bibinfo {volume} {427}},\ \bibinfo {pages} {3435} (\bibinfo {year}
  {2013})},\ \Eprint {https://arxiv.org/abs/1203.6594} {arXiv:1203.6594
  [astro-ph.CO]} \BibitemShut {NoStop}%
\bibitem [{\citenamefont {Sarkar}(2004)}]{Sarkar:2002er}%
  \BibitemOpen
  \bibfield  {author} {\bibinfo {author} {\bibfnamefont {S.}~\bibnamefont
  {Sarkar}},\ }\bibfield  {title} {\bibinfo {title} {{Measuring the baryon
  content of the universe}: {BBN vs CMB}},\ }in\ \href@noop {} {\emph {\bibinfo
  {booktitle} {{13th Rencontres de Blois on Frontiers of the Universe}}}}\
  (\bibinfo {year} {2004})\ pp.\ \bibinfo {pages} {53--63},\ \Eprint
  {https://arxiv.org/abs/astro-ph/0205116} {arXiv:astro-ph/0205116}
  \BibitemShut {NoStop}%
\bibitem [{\citenamefont {Tytler}\ \emph {et~al.}(2000)\citenamefont {Tytler},
  \citenamefont {O'Meara}, \citenamefont {Suzuki},\ and\ \citenamefont
  {Lubin}}]{TYTLER2000409}%
  \BibitemOpen
  \bibfield  {author} {\bibinfo {author} {\bibfnamefont {D.}~\bibnamefont
  {Tytler}}, \bibinfo {author} {\bibfnamefont {J.~M.}\ \bibnamefont {O'Meara}},
  \bibinfo {author} {\bibfnamefont {N.}~\bibnamefont {Suzuki}},\ and\ \bibinfo
  {author} {\bibfnamefont {D.}~\bibnamefont {Lubin}},\ }\bibfield  {title}
  {\bibinfo {title} {Deuterium and the baryonic density of the universe},\
  }\href {https://doi.org/https://doi.org/10.1016/S0370-1573(00)00032-6}
  {\bibfield  {journal} {\bibinfo  {journal} {Physics Reports}\ }\textbf
  {\bibinfo {volume} {333-334}},\ \bibinfo {pages} {409} (\bibinfo {year}
  {2000})}\BibitemShut {NoStop}%
\bibitem [{\citenamefont {Thuan}\ and\ \citenamefont
  {Izotov}(2002)}]{Thuan:2001zc}%
  \BibitemOpen
  \bibfield  {author} {\bibinfo {author} {\bibfnamefont {T.~X.}\ \bibnamefont
  {Thuan}}\ and\ \bibinfo {author} {\bibfnamefont {Y.~I.}\ \bibnamefont
  {Izotov}},\ }\bibfield  {title} {\bibinfo {title} {{The primordial helium-4
  abundance determination: systematic effects}},\ }\href
  {https://doi.org/10.1023/A:1015838716176} {\bibfield  {journal} {\bibinfo
  {journal} {Space Sci. Rev.}\ }\textbf {\bibinfo {volume} {100}},\ \bibinfo
  {pages} {263} (\bibinfo {year} {2002})},\ \Eprint
  {https://arxiv.org/abs/astro-ph/0112348} {arXiv:astro-ph/0112348}
  \BibitemShut {NoStop}%
\bibitem [{\citenamefont {Coc}\ \emph {et~al.}(2002)\citenamefont {Coc},
  \citenamefont {Vangioni-Flam}, \citenamefont {Casse},\ and\ \citenamefont
  {Rabiet}}]{Coc:2002tr}%
  \BibitemOpen
  \bibfield  {author} {\bibinfo {author} {\bibfnamefont {A.}~\bibnamefont
  {Coc}}, \bibinfo {author} {\bibfnamefont {E.}~\bibnamefont {Vangioni-Flam}},
  \bibinfo {author} {\bibfnamefont {M.}~\bibnamefont {Casse}},\ and\ \bibinfo
  {author} {\bibfnamefont {M.}~\bibnamefont {Rabiet}},\ }\bibfield  {title}
  {\bibinfo {title} {{Constraints on Omega b from nucleosynthesis of Li-7 in
  the standard big bang model}},\ }\href
  {https://doi.org/10.1103/PhysRevD.65.043510} {\bibfield  {journal} {\bibinfo
  {journal} {Phys. Rev. D}\ }\textbf {\bibinfo {volume} {65}},\ \bibinfo
  {pages} {043510} (\bibinfo {year} {2002})},\ \Eprint
  {https://arxiv.org/abs/astro-ph/0111077} {arXiv:astro-ph/0111077}
  \BibitemShut {NoStop}%
\bibitem [{\citenamefont {Sahni}\ \emph {et~al.}(2003)\citenamefont {Sahni},
  \citenamefont {Saini}, \citenamefont {Starobinsky},\ and\ \citenamefont
  {Alam}}]{Sahni:2002fz}%
  \BibitemOpen
  \bibfield  {author} {\bibinfo {author} {\bibfnamefont {V.}~\bibnamefont
  {Sahni}}, \bibinfo {author} {\bibfnamefont {T.~D.}\ \bibnamefont {Saini}},
  \bibinfo {author} {\bibfnamefont {A.~A.}\ \bibnamefont {Starobinsky}},\ and\
  \bibinfo {author} {\bibfnamefont {U.}~\bibnamefont {Alam}},\ }\bibfield
  {title} {\bibinfo {title} {{Statefinder: A New geometrical diagnostic of dark
  energy}},\ }\href {https://doi.org/10.1134/1.1574831} {\bibfield  {journal}
  {\bibinfo  {journal} {JETP Lett.}\ }\textbf {\bibinfo {volume} {77}},\
  \bibinfo {pages} {201} (\bibinfo {year} {2003})},\ \Eprint
  {https://arxiv.org/abs/astro-ph/0201498} {arXiv:astro-ph/0201498}
  \BibitemShut {NoStop}%
\bibitem [{\citenamefont {Sahni}\ \emph {et~al.}(2008)\citenamefont {Sahni},
  \citenamefont {Shafieloo},\ and\ \citenamefont {Starobinsky}}]{Sahni:2008xx}%
  \BibitemOpen
  \bibfield  {author} {\bibinfo {author} {\bibfnamefont {V.}~\bibnamefont
  {Sahni}}, \bibinfo {author} {\bibfnamefont {A.}~\bibnamefont {Shafieloo}},\
  and\ \bibinfo {author} {\bibfnamefont {A.~A.}\ \bibnamefont {Starobinsky}},\
  }\bibfield  {title} {\bibinfo {title} {{Two new diagnostics of dark
  energy}},\ }\href {https://doi.org/10.1103/PhysRevD.78.103502} {\bibfield
  {journal} {\bibinfo  {journal} {Phys. Rev. D}\ }\textbf {\bibinfo {volume}
  {78}},\ \bibinfo {pages} {103502} (\bibinfo {year} {2008})},\ \Eprint
  {https://arxiv.org/abs/0807.3548} {arXiv:0807.3548 [astro-ph]} \BibitemShut
  {NoStop}%
\bibitem [{\citenamefont {Riess}\ \emph {et~al.}(2019)\citenamefont {Riess},
  \citenamefont {Casertano}, \citenamefont {Yuan}, \citenamefont {Macri},\ and\
  \citenamefont {Scolnic}}]{Riess:2019cxk}%
  \BibitemOpen
  \bibfield  {author} {\bibinfo {author} {\bibfnamefont {A.~G.}\ \bibnamefont
  {Riess}}, \bibinfo {author} {\bibfnamefont {S.}~\bibnamefont {Casertano}},
  \bibinfo {author} {\bibfnamefont {W.}~\bibnamefont {Yuan}}, \bibinfo {author}
  {\bibfnamefont {L.~M.}\ \bibnamefont {Macri}},\ and\ \bibinfo {author}
  {\bibfnamefont {D.}~\bibnamefont {Scolnic}},\ }\bibfield  {title} {\bibinfo
  {title} {{Large Magellanic Cloud Cepheid Standards Provide a 1\% Foundation
  for the Determination of the Hubble Constant and Stronger Evidence for
  Physics beyond $\Lambda$CDM}},\ }\href
  {https://doi.org/10.3847/1538-4357/ab1422} {\bibfield  {journal} {\bibinfo
  {journal} {Astrophys. J.}\ }\textbf {\bibinfo {volume} {876}},\ \bibinfo
  {pages} {85} (\bibinfo {year} {2019})},\ \Eprint
  {https://arxiv.org/abs/1903.07603} {arXiv:1903.07603 [astro-ph.CO]}
  \BibitemShut {NoStop}%
\end{thebibliography}%
\end{document}